\documentclass[a4paper,12pt]{book}
\usepackage{amssymb}
\usepackage[italian]{babel}   
\usepackage[latin1]{inputenc} 
\usepackage{graphicx}         
\usepackage{amsmath}

\begin{document}

\begin{titlepage}
\centering
      \includegraphics[width=7.5cm]{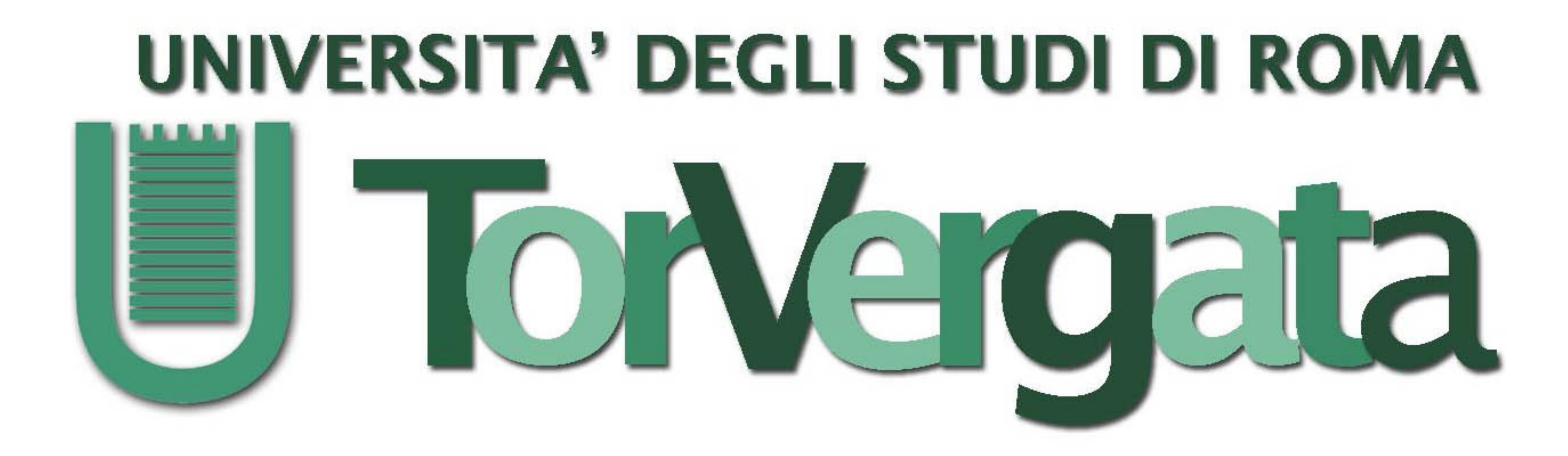}\\
     \vspace{3em}
     {\Large \textsc{Facolt\'a di Ingegneria}}\\
     \vspace{1em}
     {\normalsize Dissertazione  di Laurea in}\\
     \vspace{1pt}
     {\Large \textsc{Ingegneria Informatica}}\\
     \vspace{3em}
     \vspace{3cm}
     {\LARGE \textbf{Come, quando e quanto un mazzo di carte, e' stato ben mischiato?}}\\
     \vspace{1em}
     {\LARGE \textbf{}}\\
\vspace{10pt} \vspace{3cm}
\begin{center}
{\normalsize  Anno Accademico 2012/2013}\\
{\normalsize 26 Febbraio 2013}
\end{center}
\vspace{6pt}
  \begin{center}
    \begin{tabular}{l c c c c c c r}
    Candidato & & & & & & &  Relatore \\[2pt]
    \large{Benjamin Isac Fargion}   & & & & & & & \large{Prof. Benedetto Scoppola}\\[2pt]
    \end{tabular}
  \end{center}
\end{titlepage}

\clearpage
\newpage

 \begin{figure}
 \begin{center}
\includegraphics[width=9.5cm]{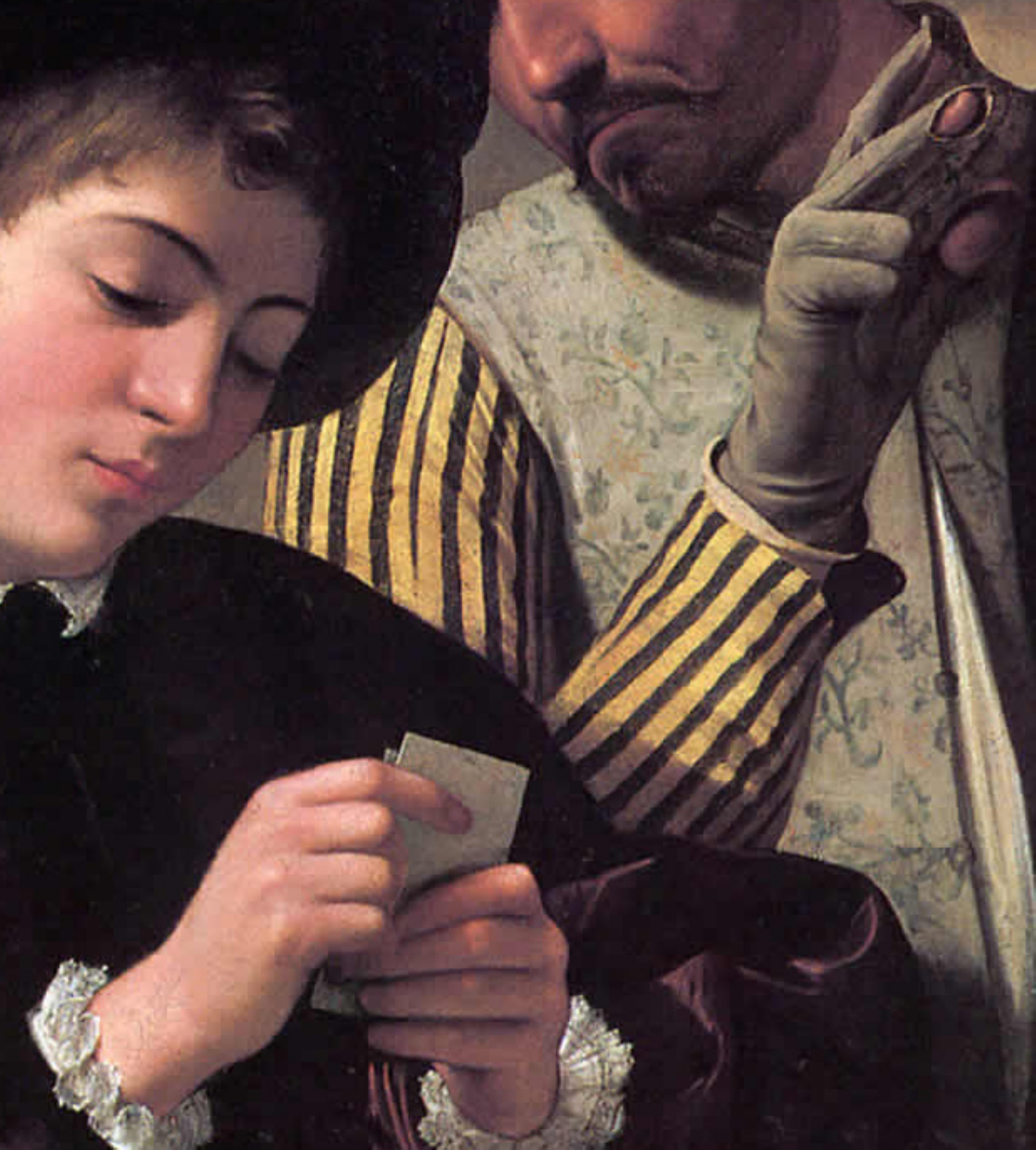}\\
\caption{Particolare del quadro del Caravaggio, il baro}
\end{center}
 \end{figure}
\vspace{2cm}
 \begin{figure}
 \begin{center}
\includegraphics[width=9.5cm]{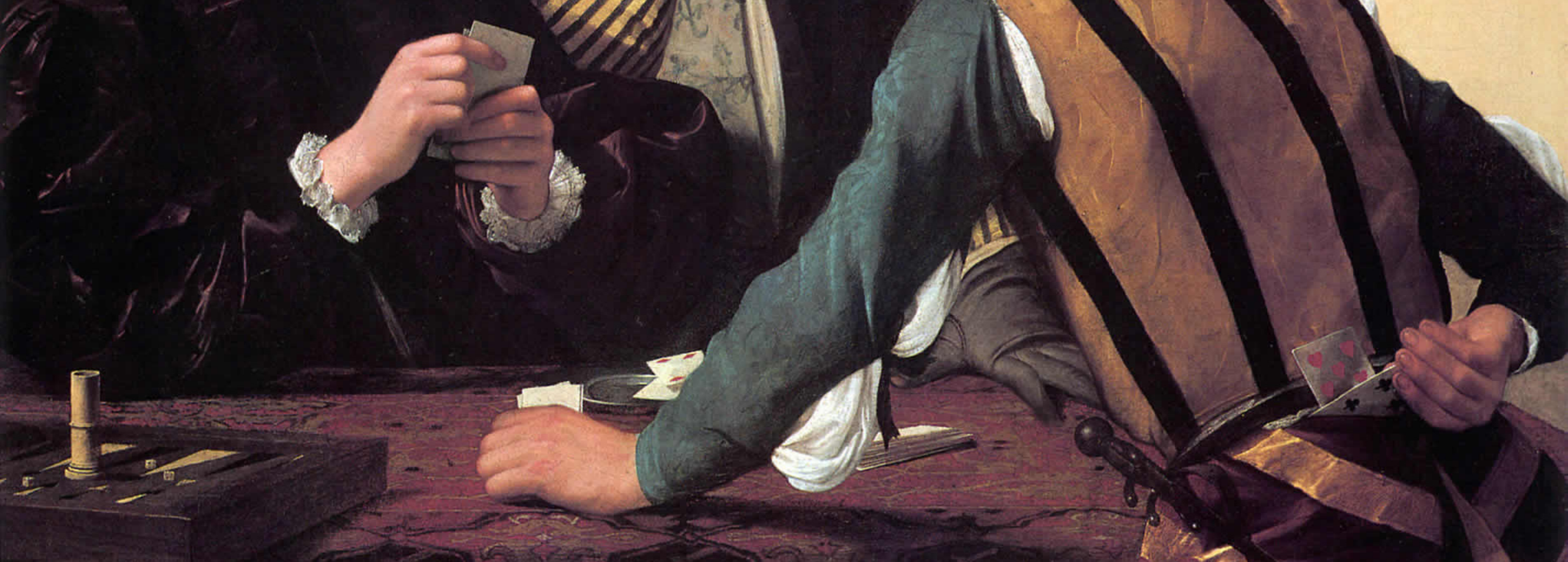}\\
\caption{Altro particolare del quadro del Caravaggio, il baro}
\end{center}
 \end{figure}
\vspace{2cm}

\begin{flushright}

\vspace{1.5cm}

\end{flushright}


\tableofcontents
\chapter{Introduzione}
In questa tesi discuteremo alcuni modelli matematici che descrivono delle pro-
cedure per stimare quanto un mazzo di carte viene mischiato. La domanda chiave
è la seguente: quanto sarà necessario mischiare un mazzo affinche' l'ordine di
questo sia abbastanza" casuale? Cio' dipenderà, come vedremo, dalla nostra
definizione di mischiare, da quanto lo si faccia e da quale grado noi si desideri
 esso sia sufficientemente" confuso. Il nostro oggetto di studio (la perdita
d'informazione) è connesso alle catene di Markov e gioca un ruolo chiave nello
sfruttare l'informazione residua nei giochi di prestigiazione. La tesi considera
alcuni tipi di shuffling noti, ove viene definita la misura del disordine tale da
permetterci di quantificare lo stato di ordine e di disordine del mazzo. Presentiamo di seguito sia alcune simulazioni numeriche per 3 e 4 carte che la descrizione analitica approssimata
che simula tale perdita di memoria nel mischiare il mazzo di carte in generale, e per mazzi di 52 carte in particolare \cite{Dia86}.

\begin{figure}[htbp]
\begin{center}
\includegraphics[scale=1.4]{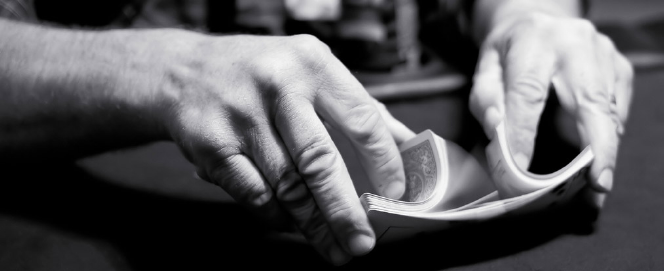}
\caption{In figura il mescolamento  di carte detto  ``riffle-shuffle'' o all'americana.
Se si mantiene una separazione esattamente a metà dei mazzetti, facendo attenzione di alternare le carte esattamente ad una ad una, si ottiene il mescolamento alla ``Faro'' (del tipo shuffle out), capace di restituire l'esatta sequenza iniziale,  dopo 8 rimescolamenti.}
\end{center}
\end{figure}

\begin{figure}[htbp]
\begin{center}
\includegraphics[scale=0.3]{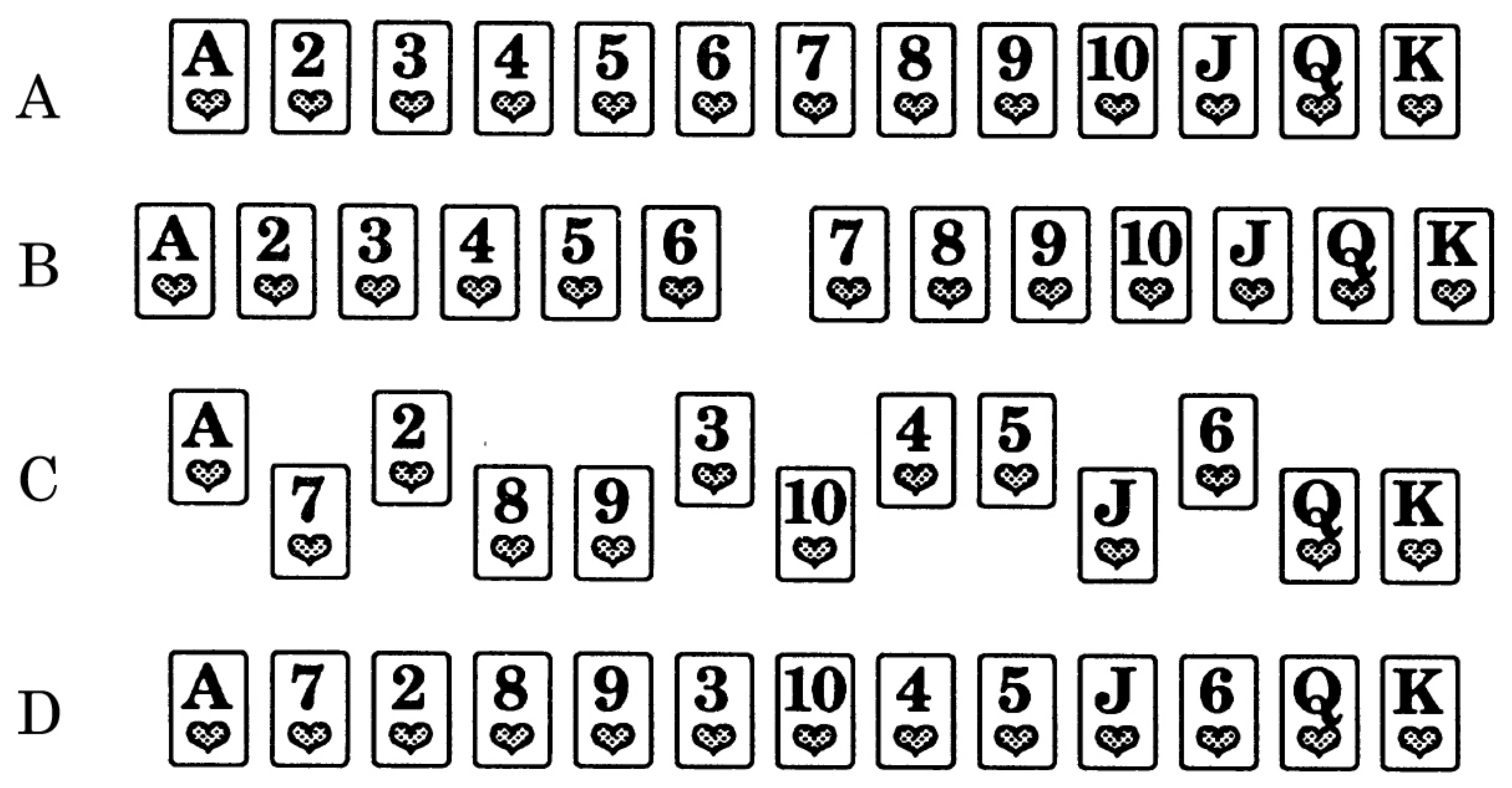}
\caption{La riga A mostra il mazzo ordinato, in riga B il mazzo originario viene diviso in due sottomazzi che vengon poi mischiati con la tecnica all'americana (riga C) ed il risultato viene mostrato in riga D.}
\end{center}
\label{dia}
\end{figure}
Nella mia tesi ho qui riproposto numericamente la procedura di mescolamento:
prima si creano due mazzi riempendo due stringhe che vengono poi frazionate come mostrato in figura\ref{dia};
in seguito s'alternano  tali sottomazzi, di lunghezza variabile  tra 1 e 5 carte ciascuno, riunendoli poi in un unico mazzo.

Nelle figure seguenti sono descritti i passi del programma e si osservano
le prime 8 mani del mazzo che va via  via perdendo l'ordine originale.
Va sottolineato che la procedura di scelta random di tutte le carte non produce affatto un simile risultato
ed è quindi errato descrivere (come in certi blog in rete) il mischiare dando a ciascuna carta delle 52 un valore a caso: tale mescolamento non corrisponde affatto al reale riffle shuffle o top-card shuffle.

Come vedremo la crescita del disordine in gran parte è dovuta alla riduzione
della sequenza in due e poi quattro o 8 e poi 16 sequenze alternate; il prestigiatore
utilizza questa informazione per individuare una qualsiasi carta inserita a caso (magicamente
 poi da indovinare) proprio dalla sua dissonante presenza nella serie delle sequenze ordinate.
Allorquando tali sequenze riducono la lunghezza del mini sequenza ad 1 sola carta,
allora non si hanno segnature ed informazioni per individuare l'ordine iniziale.
Questo avviene se le operazioni sono  di meta ogni volta, quando si dividono 52
carte per $2^{n}> 52 $  ovvero ciò avviene per  per $n > 6$. In realtà tale caso di mescolamento ideale al $50\%$ è raro (anche se tra i più probabili); esso non considera i casi asimmetrici
(piccoli e grandi mazzi, simili, al limite al top card shuffle) in cui la perdita
 di memoria (assomigliando di più al top card) è minore. Tale possibilità conduce ad un numero superiore
 (di circa 8.5 mani di shuffle, secondo il lavoro di Diaconis \cite{Dia86},\cite{Dia92}) per ottenere, in media, un buon mescolamento del mazzo  di carte come desiderato. Vedremo che questo "buon" mescolamento
 avviene quando la cosidetta distanza dall'omogeneità, in origine pari ad uno,  tende dopo numerose mani  improvvisamente a zero.

 Sono stati realizzati molti programmi per ottenere previsioni sul risultato di un' operazione di mescolamento carte. Ma il software che si può trovare in rete utilizza sempre funzioni di randomizzazione generiche che non rispecchiano la realtà della procedura fisica. Per esempio, la routine scritta in linguaggio C, riportata poco più avanti, trovata in rete al link

		$$http://www.fredosaurus.com/notes-cpp/misc/random-shuffle.html $$

effettua il mescolamento assumendo di estrarre le carte secondo una banale randomizzazione, per cui, come si può evincere dal blocco "while", che si inizia dall' istruzione 15, dopo ogni carta la successiva viene selezionata a caso fra le rimanenti del mazzo, senza tener conto della sua reale posizione fisica. Non è infatti corretto immaginare che, in una operazione di mescolamento all' americana (riffle-shuffle), dopo ogni gruppo di carte provenienti da un semi mazzo possa posizionarsi, per esempio, l' ultima presente nello stesso semi mazzo.
\begin{figure}[htbp]
\begin{center}
\includegraphics[scale=0.752]{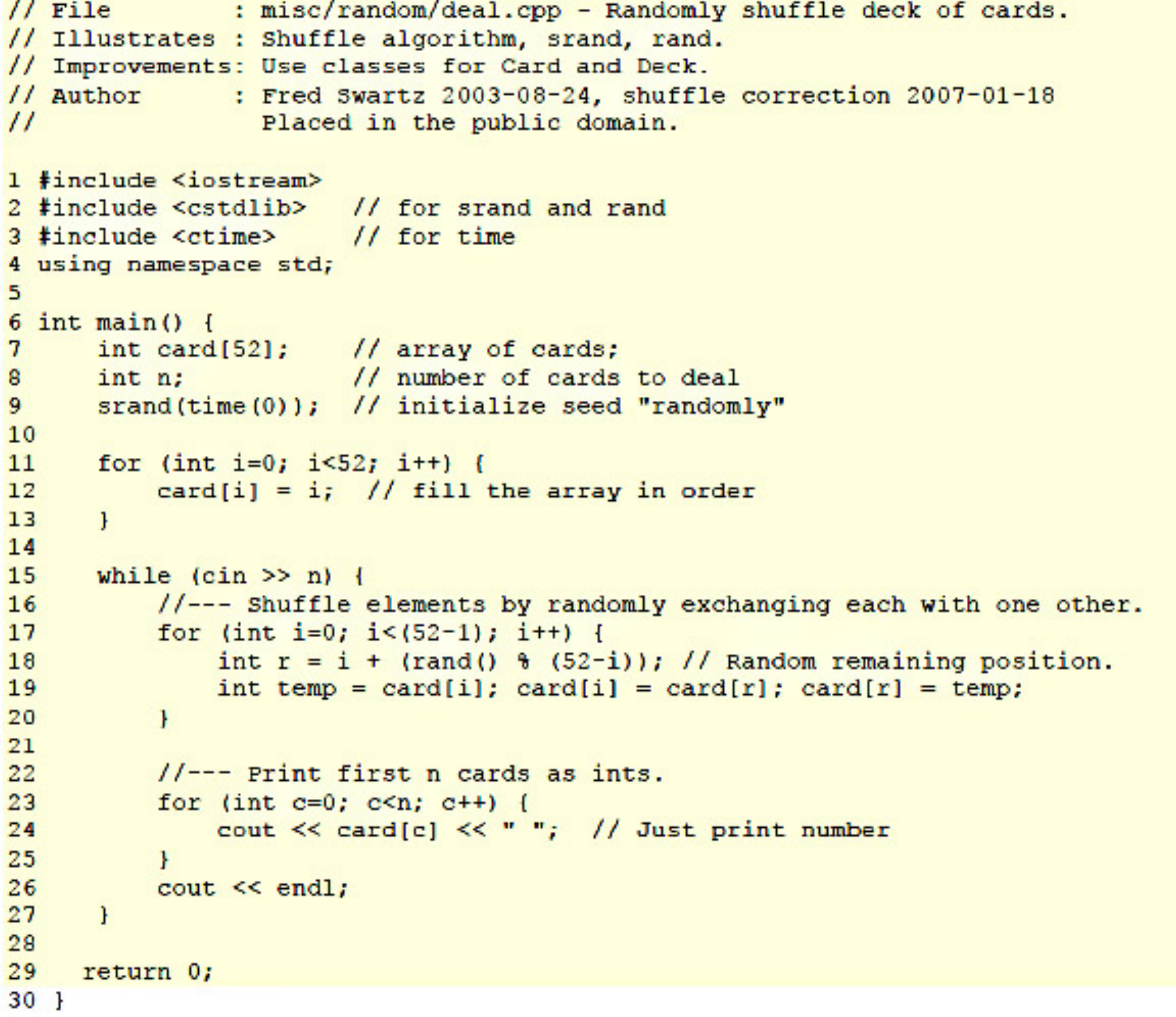}
\caption{Programma C++ in rete per un immaginario riffle shuffle }
\end{center}
\end{figure}
Di conseguenza abbiamo cercato di realizzare, con una routine scritta in Mathematica, un software che tenesse conto fisicamente della distribuzione iniziale della carte nel mazzo e fosse in grado di seguirne l'evoluzione temporale, ipotizzando che nel mescolamento all'americana le carte nei due semi mazzi si dispongano nella nuova sequenza solo se dopo ogni gruppo di n carte appartenenti a uno dei due semi mazzi si possano sovrapporre al massimo n carte (con n valore randomizzato per es. tra 1 e 5) appartenenti al secondo semimazzo.

\begin{figure}[htbp]
\begin{center}
\includegraphics[scale=0.956]{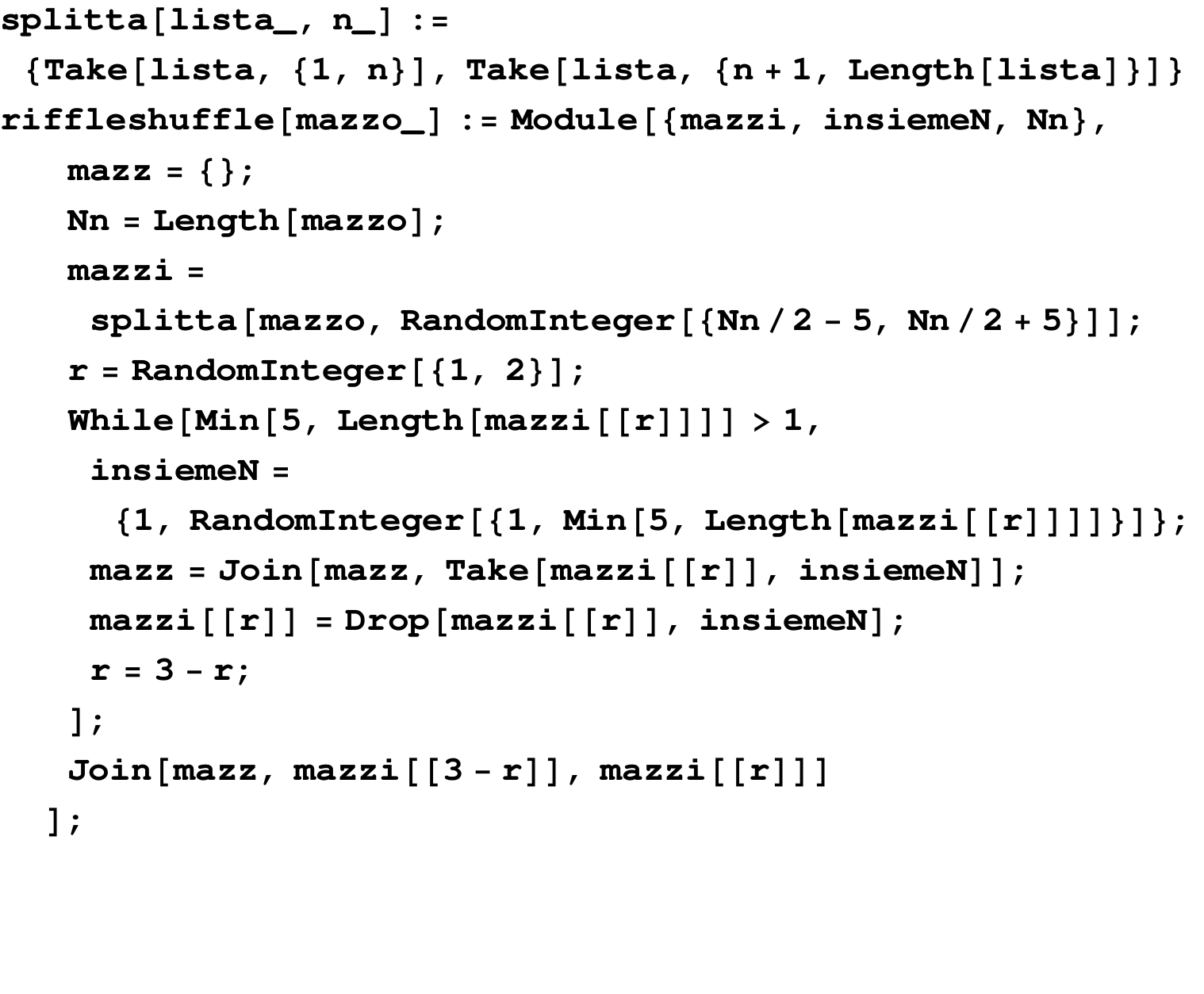}
\caption{Questo programma in Mathematica si occupa di dividere il mazzo di
52 carte, analogamente a quanto sopra descritto, seguendo una sequenza di operazioni: si
separano le 52 carte in un gruppo di n carte pari a $26 +/- 5$  e le rimanenti$26 -/+ 5$ dall'altra parte. Quindi si selezionano delle frazioni di un mazzo (tra 1 e 5 carte) a gruppi alternati (a pettine) con
il secondo mazzo; così si simula il mix all'americana o il riffle shuffle. Questa prima
immagine ne descrive un mazzo originale ordinato.  }
\end{center}
\end{figure}
La prima funzione (Splitta) effettua la separazione del mazzo iniziale in 2 semi-mazzi: se si pone $n=26$, si ottengono due semi-mazzi identici di 26 carte l'uno e se si procedesse in modo esatto alternato si otterrebbe
il mescolamento Faro, di cui accennerò oltre. Qui considero il caso random in cui i due mazzi non siano
necessariamente identici e non debbano quindi essere mescolati a Faro.
La funzione riffleshuffle preleva da ciascun semi mazzo sequenze di al massimo 5 carte.
Il blocco While preleva alternativamente dai 2 semi-mazzi  alcuni mazzetti di al massimo 5 carte, fintanto che i semimazzi originali contengono ancora delle carte.
La variabile r ( e la sua congiunta, $3-r$) assume un valore random tra 1 e 2 per indicare da quale dei 2 semi-mazzi ha inizio il mescolamento, in modo che la funzione Join restituisca il risultato del mescolamento dei 2 semimazzi.
\begin{figure}[htbp]
\begin{center}
\includegraphics[scale=1.16]{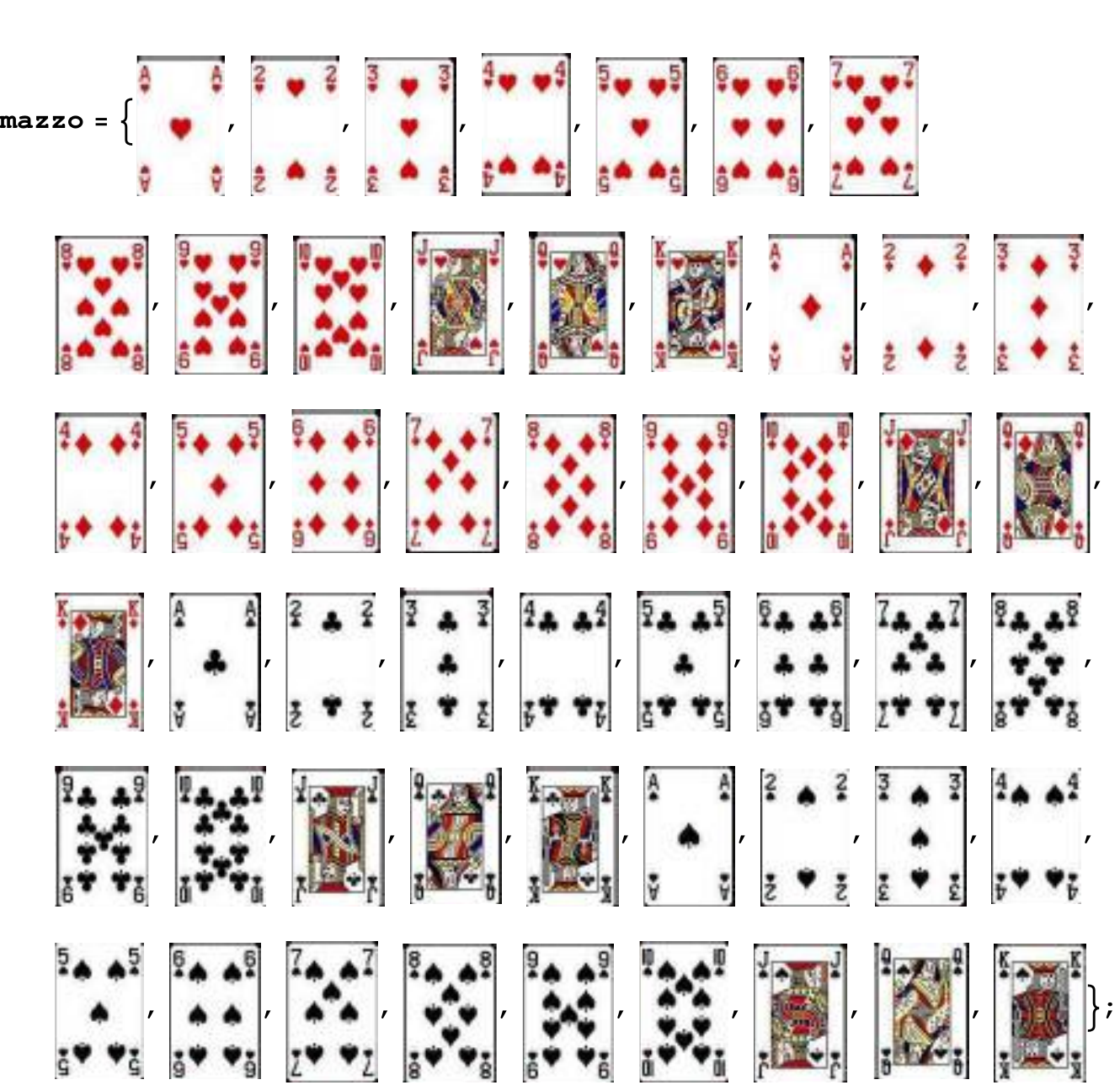}
\caption{Definizione del mazzo ordinato; esso è l'oggetto su cui gli operatori di programma descritti dalle routine precedenti agiscono per il mescolamento a pettine (riffle shuffle) all'americana. }
\end{center}
\end{figure}

\begin{figure}[htbp]
\begin{center}
\includegraphics[scale=0.9]{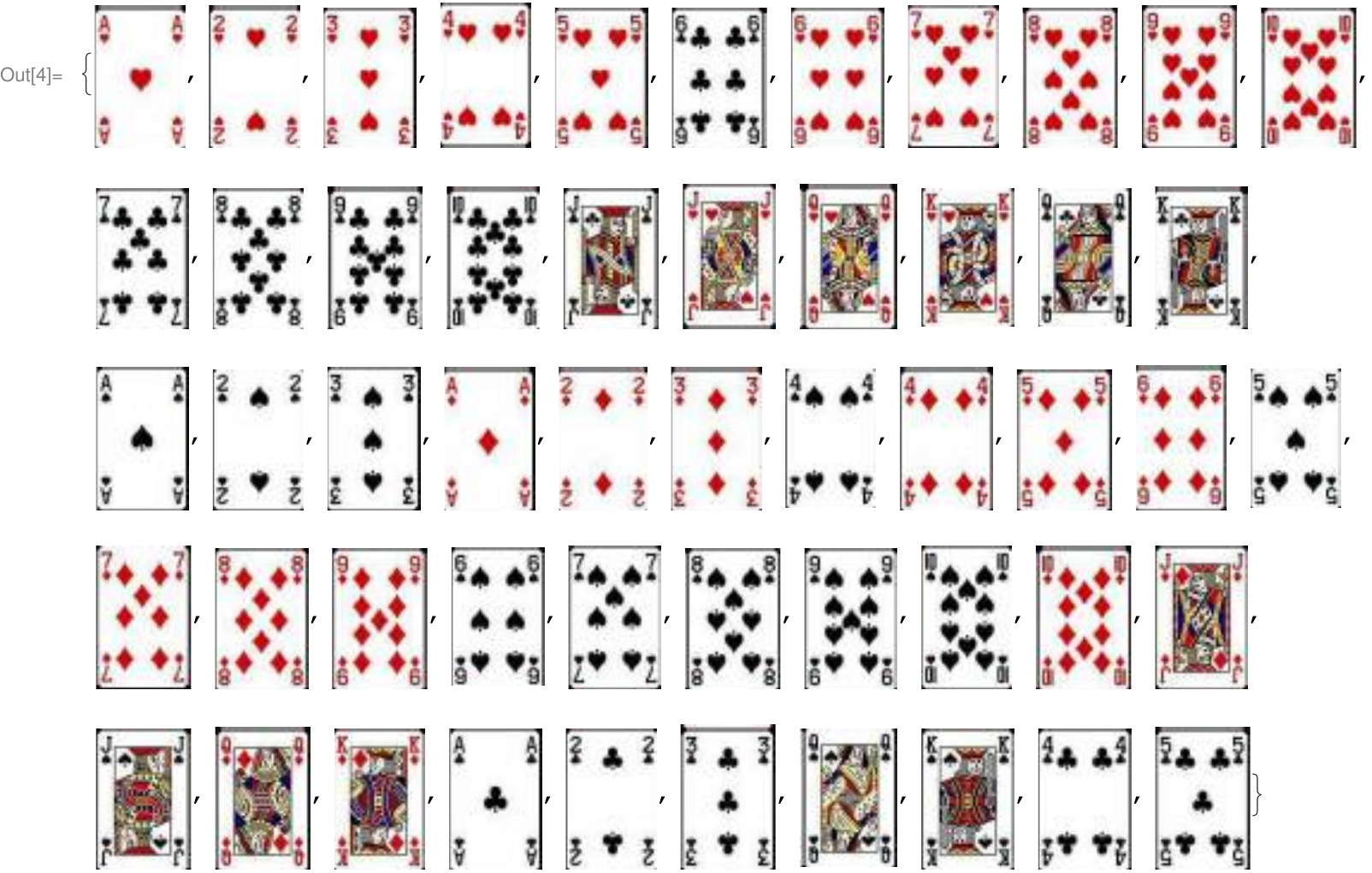}
\caption{Ecco una disposizione ottenuta alla seconda mano ovvero dopo un primo mescolamento riffle  }
\end{center}
\end{figure}

\begin{figure}[htbp]
\begin{center}
\includegraphics[scale=0.55]{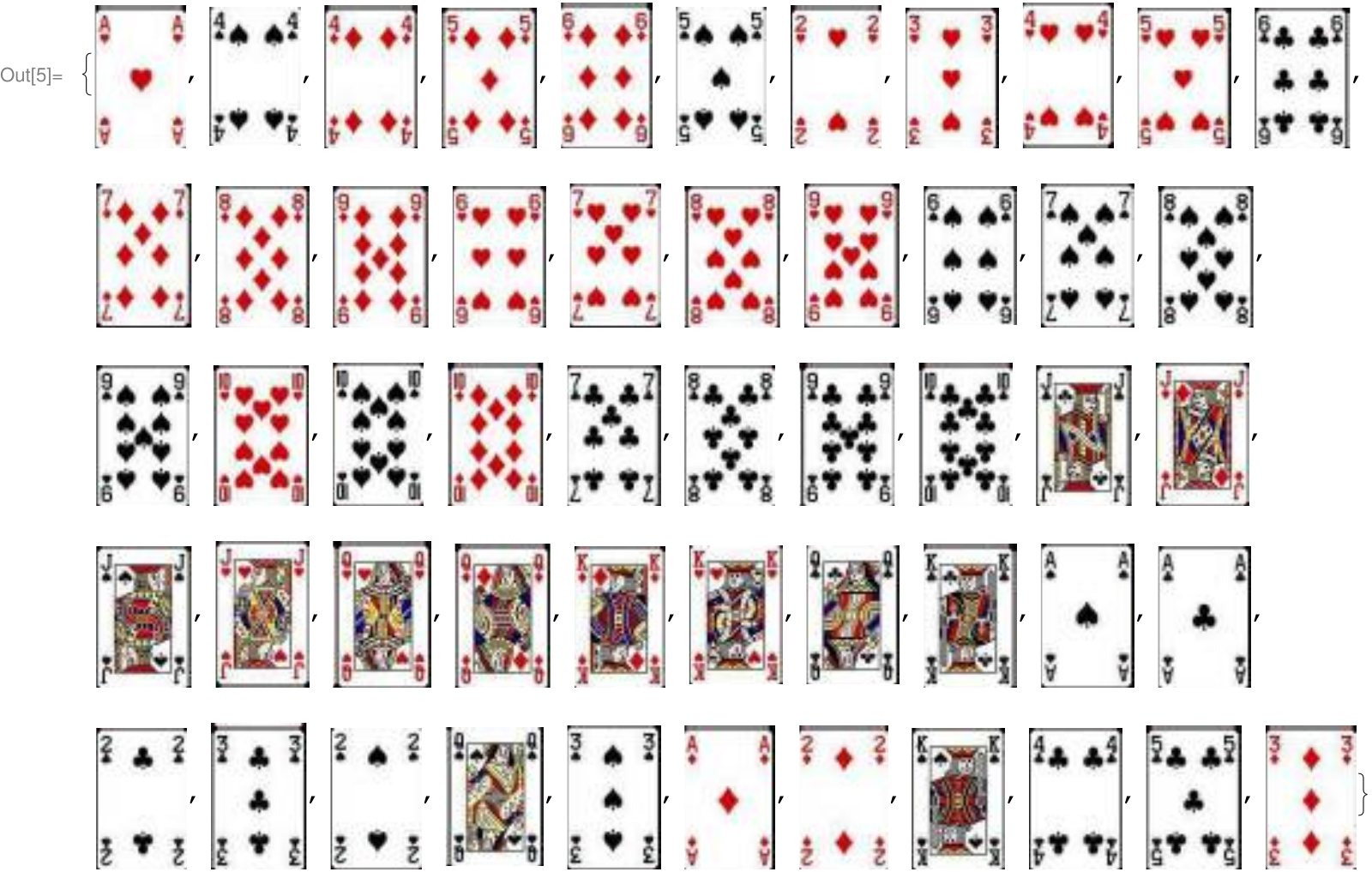}
\caption{Come sopra per una terza  mano ovvero dopo un secondo mescolamento riffle  shuffle }
\includegraphics[scale=0.55]{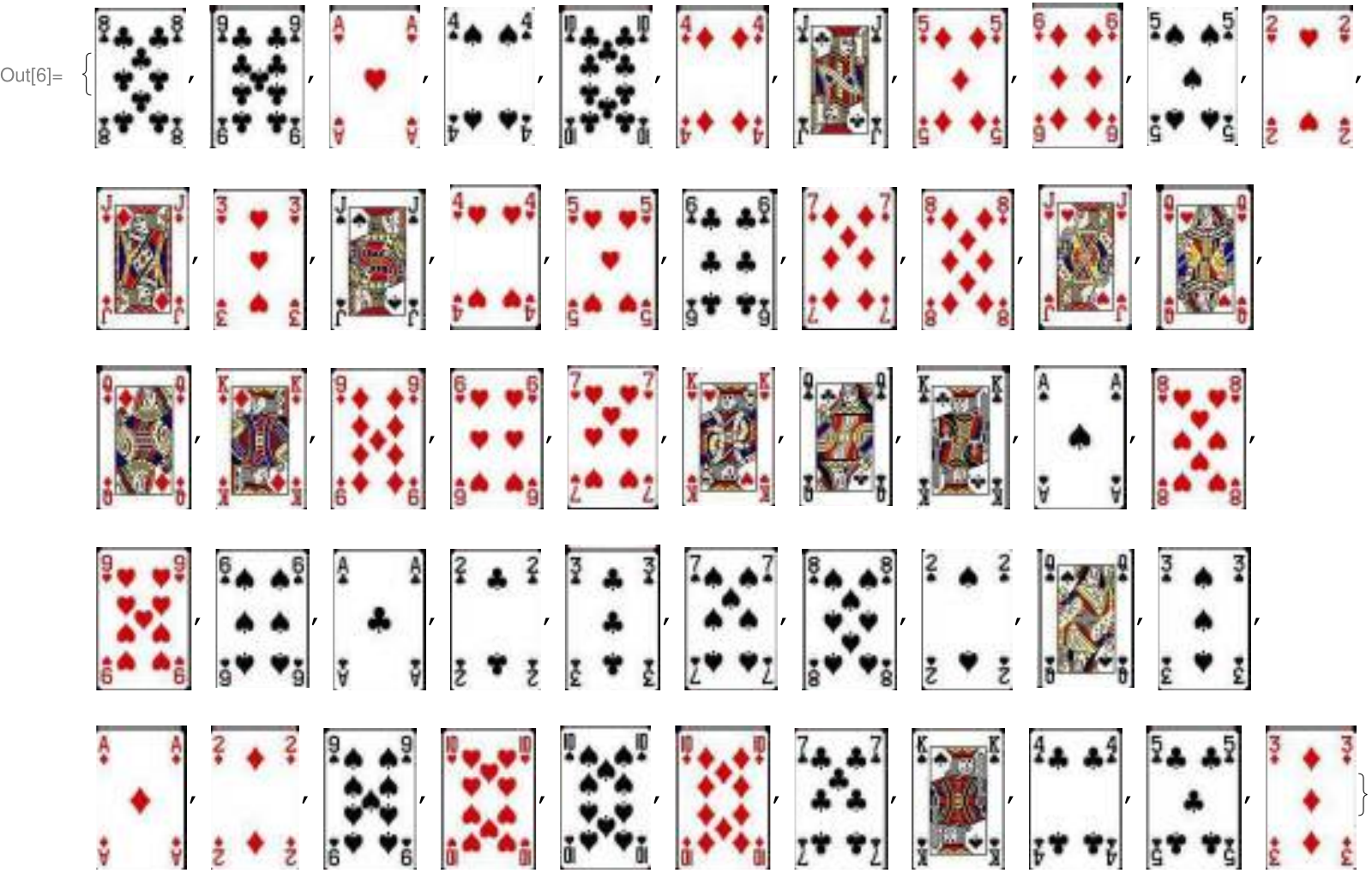}
\caption{Come sopra la sequenza per una quarta  mano ovvero dopo un terzo mescolamento riffle  shuffle  }
\end{center}
\end{figure}

\begin{figure}[htbp]
\begin{center}
\includegraphics[scale=0.55]{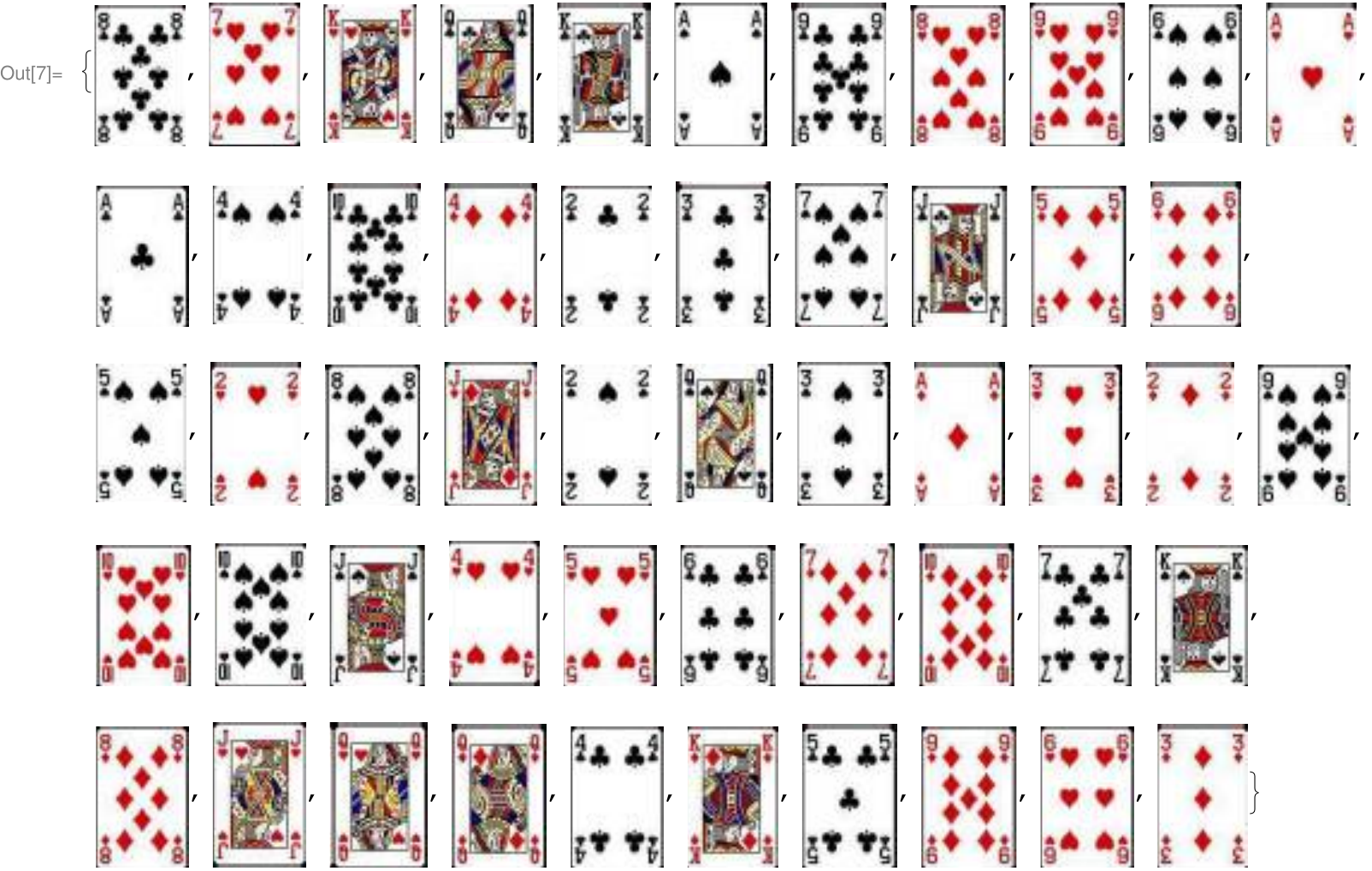}
\caption{Come sopra per una quinta  mano ovvero dopo un quarto mescolamento riffle  shuffle   }
\includegraphics[scale=0.55]{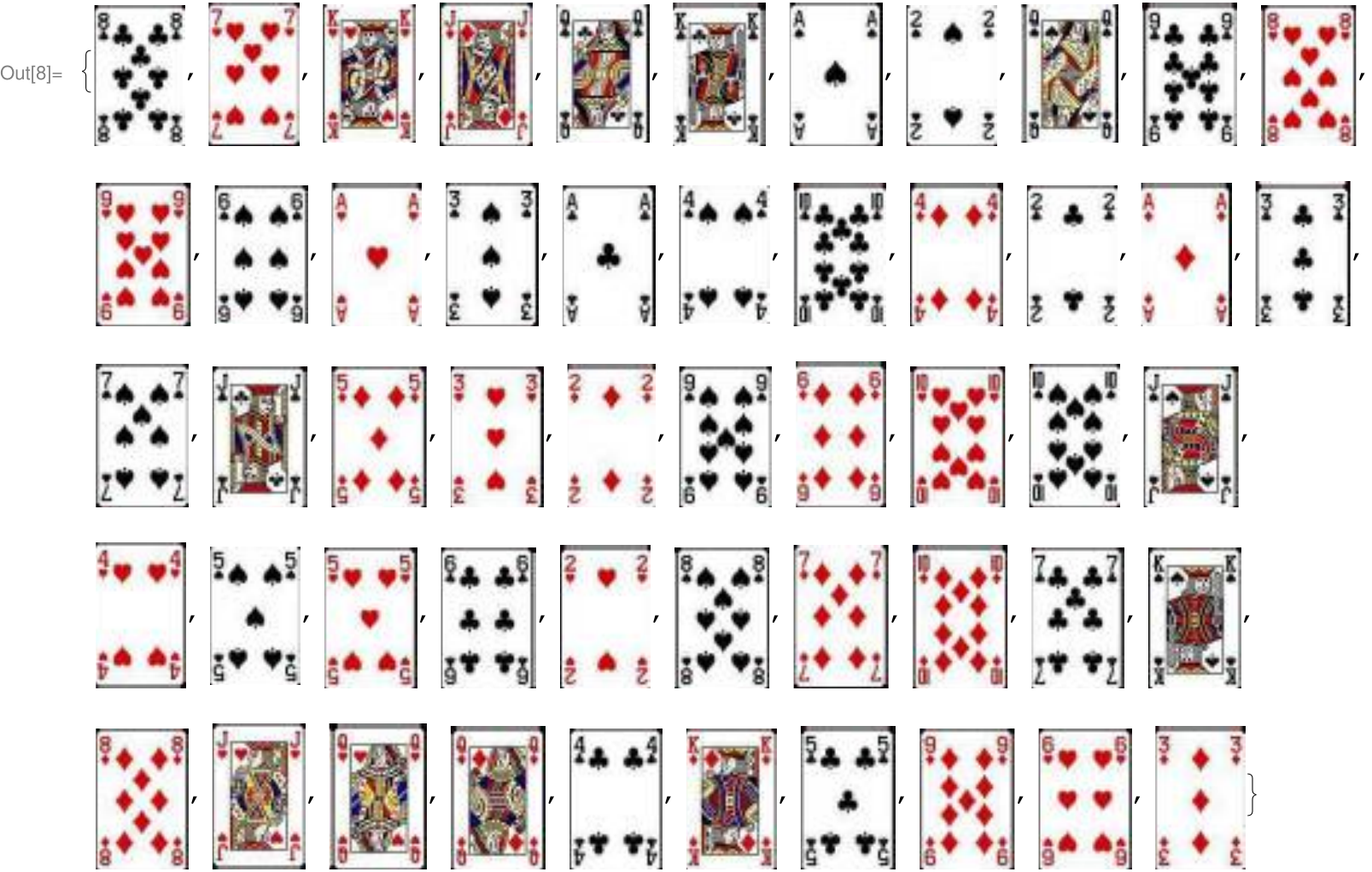}
\caption{Come sopra per una sesta mano ovvero dopo un quinto  mescolamento riffle  shuffle  }
\end{center}
\end{figure}

\begin{figure}[htbp]
\begin{center}
\includegraphics[scale=0.55]{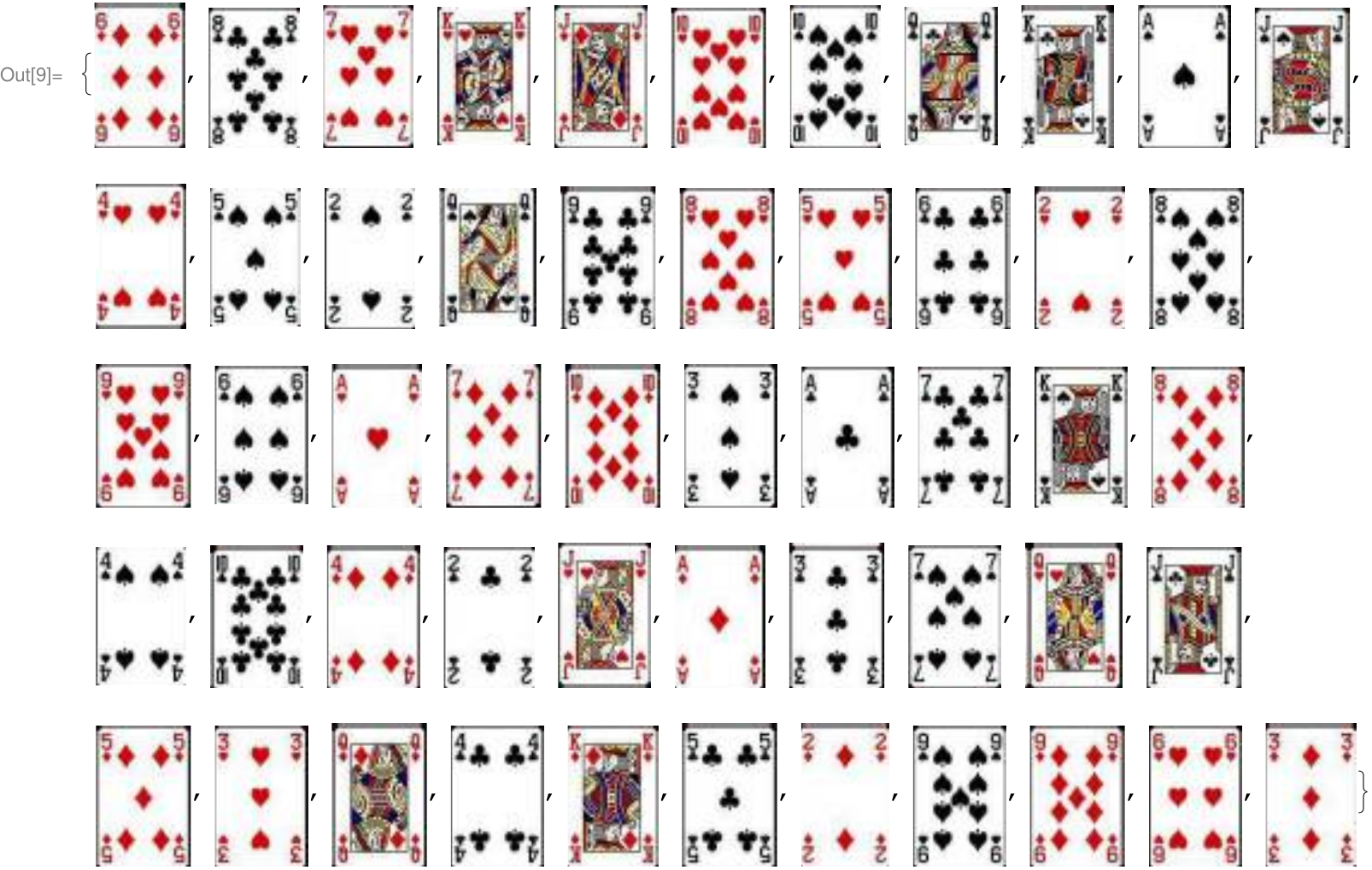}
\caption{Come sopra per una settimana   mano ovvero dopo un sesto mescolamento riffle  shuffle  }
\includegraphics[scale=0.55]{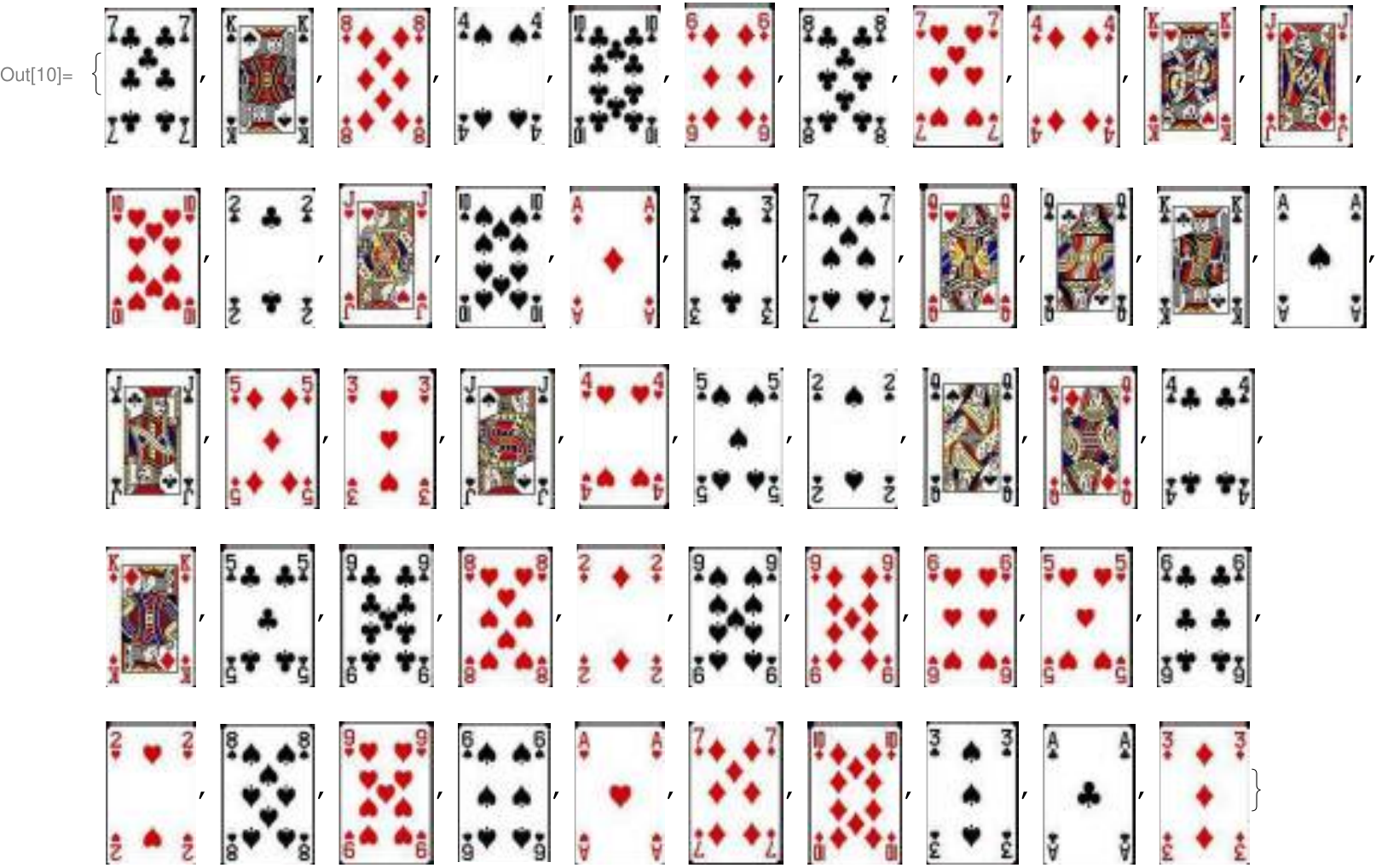}
\caption{Come sopra per una ottava  mano ovvero dopo un settimo mescolamento riffle  shuffle  }
\end{center}
\end{figure}

\chapter{Quanti modi per mischiare un mazzo di carte?}
 Ricordiamo alcune delle principali metodologie di mescolamento delle carte:
 \begin{itemize}
 \item \textbf{Top Card Shuffle} Si trasporta e si inserisce la carta più alta (top card) all'interno del mazzo in modo random; questo  è il metodo più semplice da  me considerato, ma, di solito anche il più lento per ottenere un definitivo mescolamento.
 \item \textbf{Riffle Shuffle (all'americana)} Tale metodo di mescolamento delle carte, tra i più popolari, già descritto nella parte introduttiva, produce un'alternanza casuale di carte prese da due sottomazzi in sequenza alternata con procedura random. Questo sarà anche oggetto del capitolo successivo.
\item \textbf{Stripping or Overhand} in questa tecnica di mescolamento, si separa dal mazzo integro un gruppo di carte e si posiziona sopra al mazzo stesso.
\item \textbf{Hindu shuffle} metodo usato maggiormente in Asia, simile allo Stripping, sopra indicato.
\item \textbf{Corgi, Chemmy, Irish o Wash shuffle} metodo di mescolamento dove le carte sono disposte tutte a faccia in giù e  poi vengono mescolate su una grande superficie.
\item \textbf{Mongean shuffle} dove le carte sono ordinate prima in serie $1, 2, 3, \ldots, 2n$ e vengono scelte  in alternanza  in serie pari-dispari decrescente-crescente ovvero $2n-2, 2n-4, \ldots,4, 2, 1, 3, 2n-1$
 (Dodici mischiate Mongean consecutive restituiscono l'ordine primordiale del mazzo di 52 carte).
 \item \textbf{Weave e Faro shuffle} in cui si divide il mazzo di 52 carte in due esatte metà  di 26, poi alternati con cura in un mix di carte alternate ad una a una. Otto mix riportano allo stadio ordinato originale.
 \end{itemize}

 Infine esistono modelli digitali di mix o randomizzazione che però pur essendo
automatici hanno algoritmi numerici in principio riproducibili e prevedibili. Uno
dei processi elementari, top card shuffle, o una carta alla volta, mette la carta
in alto all'interno del mazzo casualmente.
Questa procedura di mescolamento ed il riffle shuffle (all'americana) sono oggetto della presente
tesi. Nel sottocapitolo successivo verrà considerato brevemente il caso Faro che è periodico ed è
perfettamente prevedibile.

\section{Faro Shuffling}
Per effettuare il Faro shuffling è necessario dividere il mazzo in due metà eguali e premere queste l'una contro l'altra affinché le carte dei due mazzetti cominciano a compenetrarsi. Se fatto a regola d'arte tale mescolamento produrrà due mezzi mazzi da 26 carte ciascuno che uniti formeranno un mazzo composto alternativamente da una carta dell'uno ed una dell'altro. Tale metodo ovviamente non randomizza il mazzo finale che conserverà quindi un alto grado d'informazione riproducendo semplicemente l'ordine alternato delle due metà di partenza.
Riproducendo per 8 volte (out of shuffle) tale mischiamento riporterà il mazzo
alla configurazione iniziale. Questo si può vedere semplicemente seguendo ad
esempio il destino di una carta particolare; sia essa per esempio la carta al 7imo
posto: seguendo la prima mischiata viene raddoppiata la posizione di ogni carta
e si avrà dunque $7\rightarrow 14$; alla seconda mano pertanto s'avrà  $14\rightarrow 28$ e alla terza  $28\rightarrow 56$; naturalmente il mazzo ha solo 52 carte ed è dunque necessario ritornare entro questo limite, ovvero  bisogna collocarsi entro un  ``modulo 51'' (se la prima resta all'esterno (out-shuffle caso)); perciò  $56-51=5$;
 continuando così s'avrà $5 \rightarrow 10$, quindi  $10 \rightarrow 20$, ed ancora $20 \rightarrow 40$ e finalmente alla settima mano $40\rightarrow 80$, ovvero grazie al modulo 51, $80-51= 29$ quindi all'ultima ottava mano si ritorna alla posizione iniziale: $29\rightarrow 58 - 51= 7$. Data l'arbitrarietà della scelta della carta (7) da seguire, fatta all'inizio, questa prova varrà per ogni carta del mazzo.
In pratica questo tipo di mescolamento mantiene la completa informazione della sequenza originale durante ogni processo di mescolamento.
  Qui di seguito osserviamo come nel out-shuffling Faro (in cui si inseriscono all'interno del mazzo  la parte superiore  tenuta dalla mano destra, ovvero la prima e l'ultima carta  restano nella lora posizione originale
  per es: 12345678$--->$ 15263748); le sequenze delle carte ordinate sono come in figure sotto indicate per 8 mescolamenti. Nel caso in-shuffle, qui non considerato, la sequenza diventa invece: (12345678$--->$51627384) e la ripetitività dura ben 26  mani per vederne una forma poi riflessa o capovolta, ma ordinata  in ben 52 mani per ritornare all'originale sequenza del mazzo.
\begin{figure}[htbp]
\begin{center}
\includegraphics[scale=0.15]{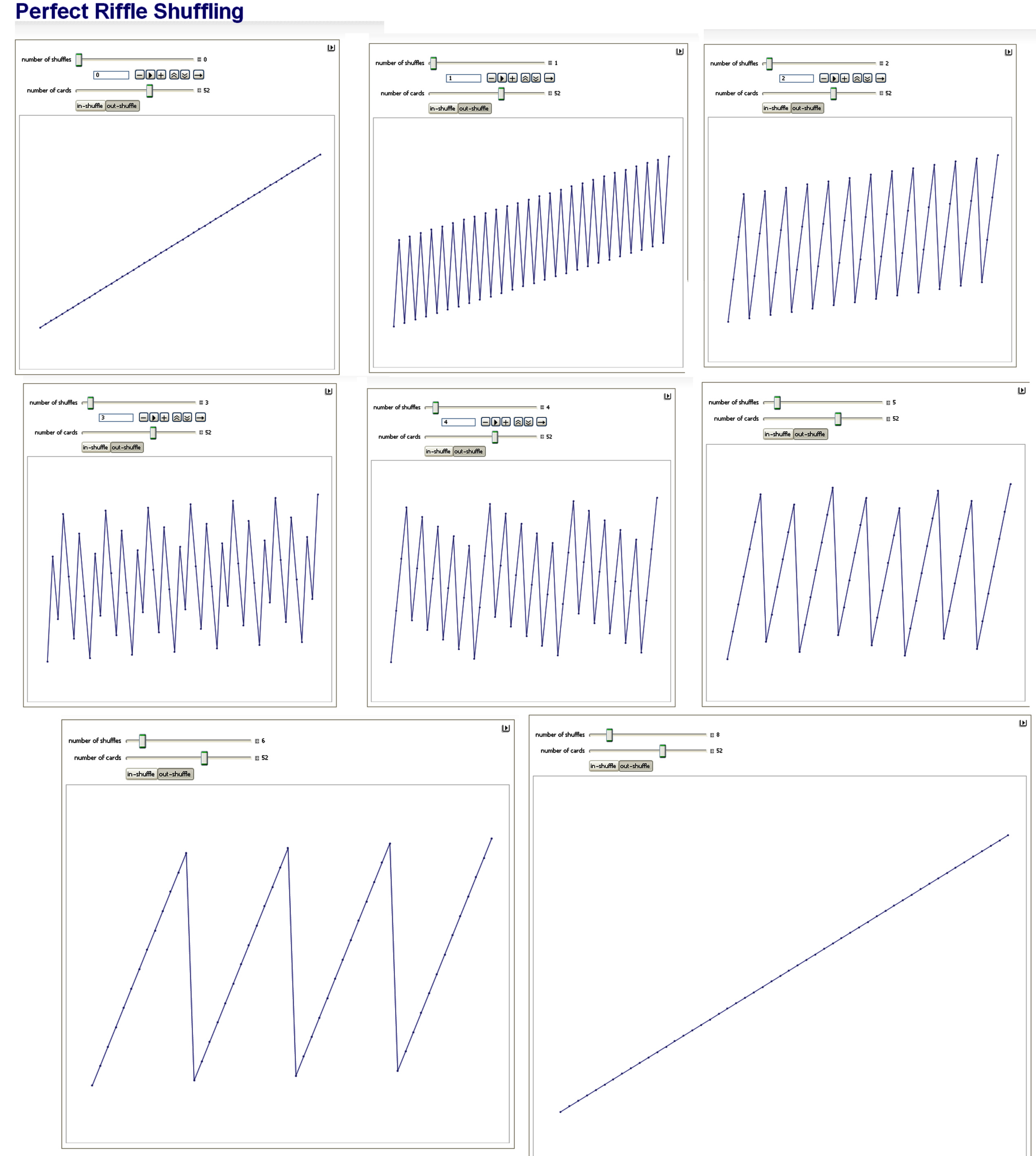}
\caption{Le evoluzioni di sequenza delle carte in un mescolamento out-shuffle Faro. Dopo 8 mescolamenti si riottiene l'ordine iniziale. }
\end{center}
\end{figure}

\chapter{Mischiare uno alla volta o a gruppo}
Consideriamo alcuni esempi di mescolamento, le  probabilità connesse e le distanza dall'uniformità ottenute per ogni caso ad una data procedura di shuffle (mescolamento del mazzo). Notiamo che le possibili permutazioni di 52 carte sono 52!, che è un numero molto grande :esso corrisponde a circa $8\cdot10^{67}$ sequenze; si prenda a paragone il numero di atomi presenti nell'universo, stimabile
moltiplicando il numero delle galassie ($10^{11}$) con il numero di stelle in ogni galassia ($10^{11}$) ed ancora per la massa tipica di una stella $4\cdot10^{33}$ gr., ed infine per il numero d'Avogadro
$6\cdot10^{23}$ ottenendo come ordine di grandezza $10^{11}*10^{11}*10^{33}*10^{24} = 10^{79}$. Questo spiega le difficoltà o meglio la impossibilità nel tener conto di ogni variabile durante il mescolamento.
 Come vedremo inoltre, al fine di descrivere le possibili sequenze dobbiamo ricorrere a matrici di probabilità di trasformazione le quali assumono dimensioni tipiche di $n!^2 = 10^{136}$.
Tale enorme entità matriciale non può essere neanche in linea di principio posta in un calcolatore essendo di una magnitudine molto maggiore di tutti gli atomi dell'universo ed anche maggiore di tutti i fotoni presenti nel nostro cosmo. Semplicemente non abbiamo spazio (e tempo) per immagazzinare le informazioni derivanti da tale calcolo matriciale.
Per tale motivo nelle normali simulazioni si usa un mazzo ridotto di tre carte (6 permutazioni) e conseguentemente una matrice 6x6. Il passo successivo è quello d'usare vettori descrivendo le configurazioni di 4 carte e le matrici di mescolamento di 24x24, ovvero  di 576 elementi. Questo test analitico che non appare in bibliografia a me nota, e' centrale in questo studio, ed è stato realizzato nella mia tesi e verrà discusso nel capitolo successivo e nelle conclusioni, sia per top-card che per riffle shuffle.

Prima di discutere altri esempi  ci sorprende che un'analisi matematica di questa domanda apparentemente innocua è collegata a una vasta gamma di questioni matematiche, in settori come algebre di Lie, omologia Hochschild, e passeggiate aleatorie su grafi, cui accenniamo solo parzialmente nel testo.

Adesso per cominciare consideriamo un mazzo di $n$ carte semplicemente etichettate con i numeri da 1 a $n$. Inizialmente le carte sono nel loro ordine naturale e noi le mischieremo
con un metodo chiamato per primo esempio come ``top-in-at-random'' shuffle. Questo tipo di shuffle consiste semplicemente nel prendere la prima carta del mazzo e riporla all'interno del mezzo  in modo casuale. Risulta evidente che tale modalità è molto più lenta del riffle shuffle (vediamo però che per N=3 avviene uno strano sorpasso nella procedura di mix rispetto al riffle shuffle); mischiando 4 carte o più  ciò non avviene: riffle  è sempre più rapido del  ``top-in-at-random'' .

Per fare un esempio semplice prendiamo tre sole carte 1, 2 e 3. Quantifichiamo le probabilità che si abbia una certa sequenza attraverso il nostro top-shuffle: all'inizio avremo ovviamente la distribuzione di probabilità (1,0,0,0,0,0) in quanto le carte sono in posizione originale.

 Cerchiamo di essere più quantitativi a proposito di questo riordino dopo una mano di mescolamento. L'insieme di tutte le permutazioni della selezione saranno indicate con $S_{n}$, che viene spesso chiamato il gruppo simmetrico di $S_{n}$ lettere. Esistono ben $n!$ possibili ordinamenti. Se si indica con $\pi$ un ordinamento particolare del mazzo, indicheremo la probabilità che esso diventi $\bar{\pi}$  dopo un shuffle grazie all'operatore Q ($\bar{\pi}$).

Si consideri l'esempio precedente in cui n = 3. In questo shuffle, togliamo la prima carta dal mazzo e la inseriamo ovunque (anche dove stava), con uguale probabilità, in una qualsiasi delle n posizioni disponibili.
Non tutte le 6 configurazioni saranno però  possibili ma solo 3 potranno avvenire dopo la prima mano:

\begin{figure}[htbp]
\begin{center}
\includegraphics[scale=0.5]{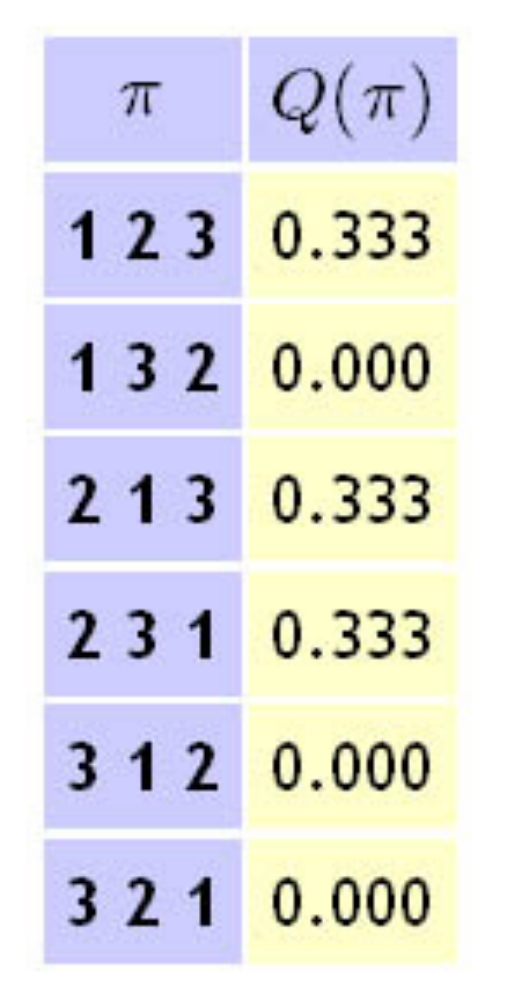}
\caption{}
\end{center}
\end{figure}
Esse (tre)  sono ugualmente probabili (un terzo ciascuna).
Questo mostra che tre dei sei possibili ordinamenti sono equi-probabili, e gli altri tre non sono affatto possibili. Così mischiamo il mazzo di nuovo ... e di nuovo ancora. I possibili ordinamenti della selezione sono riportati di seguito dopo 1, 2, e 3 shuffle. Per comodità, le carte sono qui rappresentate da colori piuttosto che essere numerati per indicarne la collocazione. Le evoluzioni di probabilità crescono come alberi
di canali all'inizio ben valutabili a mano, e poi via via meglio prevedibili dalle moltiplicazioni delle matrici (6 x 6) Q, matrici alla prima o quadrata.., che sono presentate di seguito.
\begin{figure}[htbp]
\begin{center}
\includegraphics[scale=0.5]{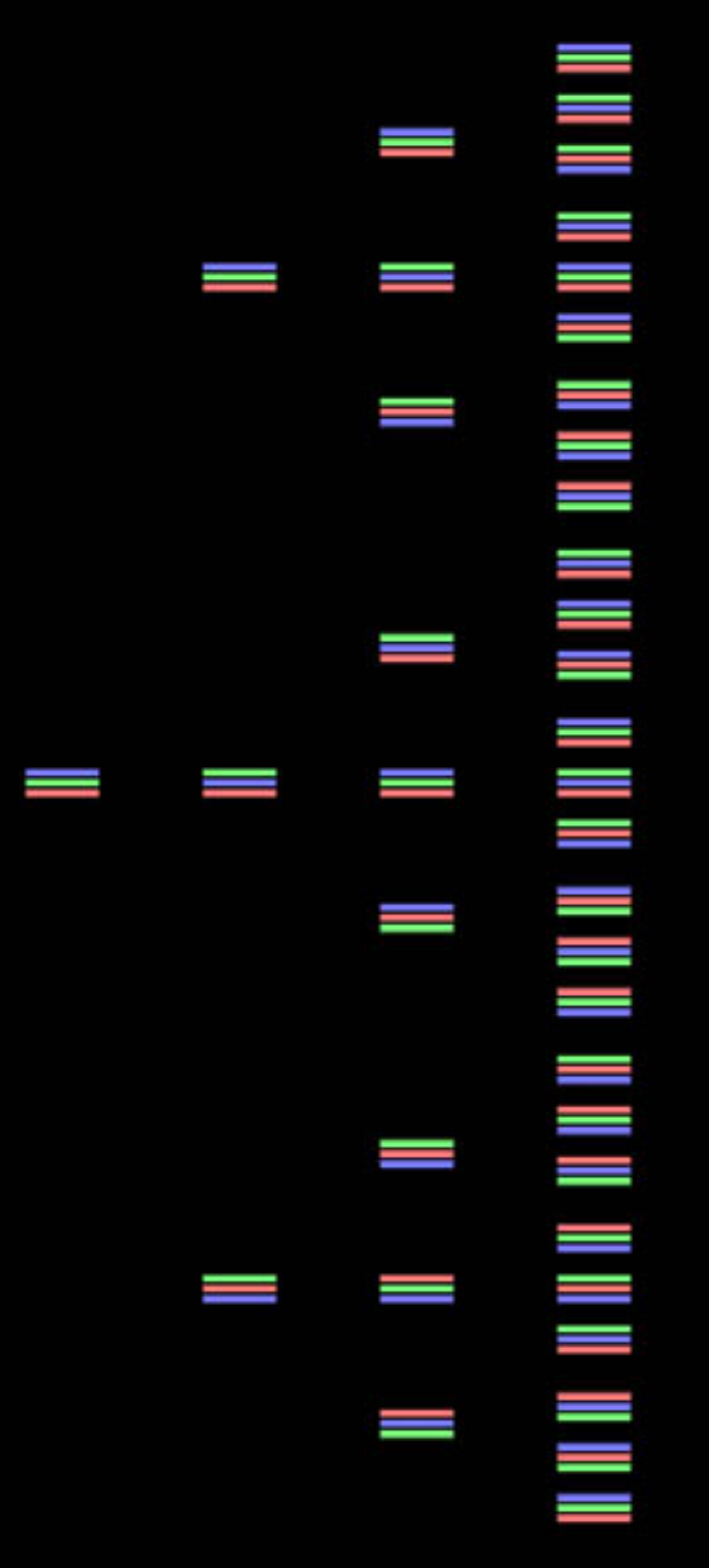}
\caption{}
\end{center}
\end{figure}
 Con $Q^k$, si intende  la procedura (ovvero l'operatore o la matrice) che calcola quale sia la probabilità che la configurazione  originale $\pi$ sia ancora quella successiva presente dopo k shuffles. Un po di calcoli elementari mostrano che (per il presente caso a 3 carte top card shuffle) vedi \ref{Fig4-Probability.pdf}.

\begin{figure}[htbp]
\begin{center}
\includegraphics[scale=0.5]{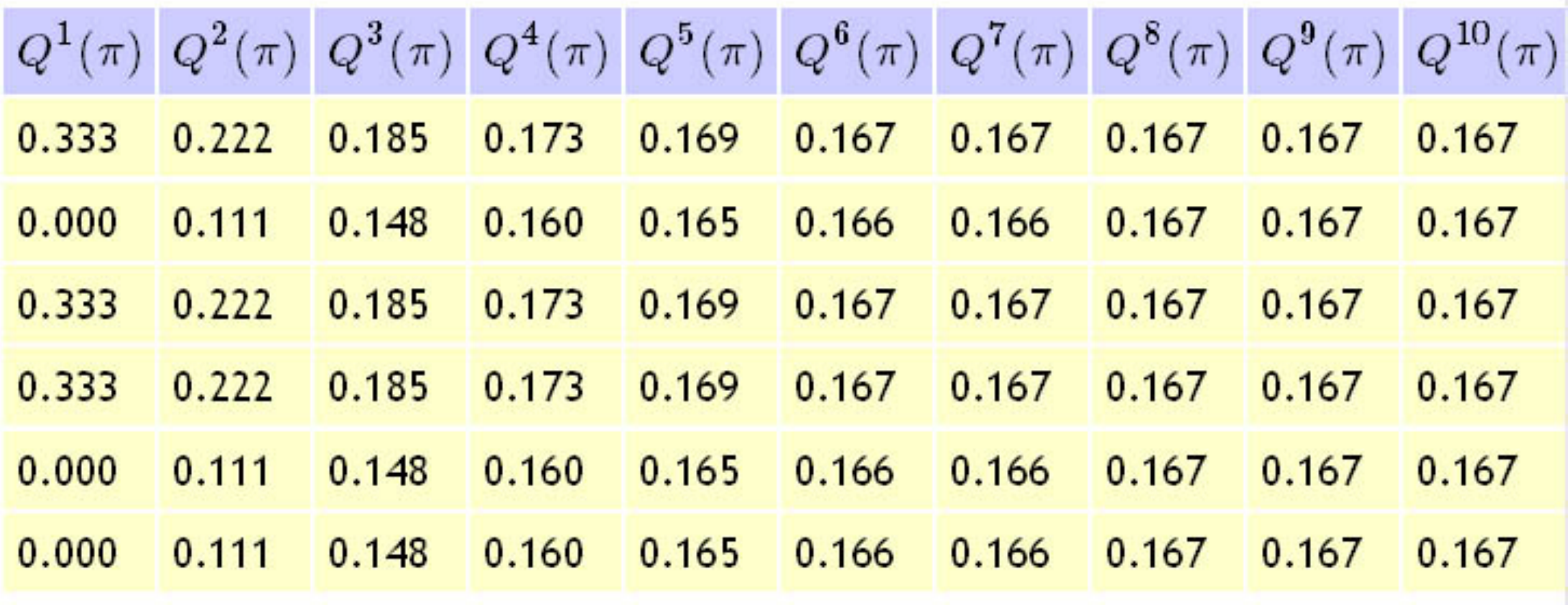}
\caption{Sequenze di probabilità delle configurazioni per 3 carte}
\label{Fig4-Probability.pdf}
\end{center}
\end{figure}
Si noti che proseguendo a mischiare, le carte tendono ad uniformarsi o meglio ad aver ugual probabilita' di accadere. Con U($\pi$),indichiamo il vettore di densità di probabilità in cui la configurazione  $\pi$ è ugualmente probabile; perciò  dopo molti mescolamenti U($\pi$)=1/n!  I risultati nella tavola, vedi  \ref{Fig4-Probability.pdf}.
 ,implicano che:
$Q^k \to U \quad\hbox{as}\quad k\to\infty.$

Per rendere questa affermazione matematicamente più chiara, e per facilitare un'analisi quantitativa del rimescolamento, abbiamo bisogno di introdurre una nozione di distanza tra le densità di probabilità. Quindi, supponiamo che $Q^{1}$ e $Q^{2}$ siano due densità di probabilità sul gruppo delle configurazioni $S_{n}$. Primo, si definisce la differenza $Q^{1}$ e $Q^{2}$ per A, un sottoinsieme di $S_{n}$, sommando le differenze su tutti gli elementi di A:

$$\|Q_1(A)-Q_2(A)\| = \frac{1}{2}\sum_{\pi\in A}|Q_1(\pi) - Q_2(\pi)|$$
Allora la distanza tra $Q_{1}$ e $Q_{2}$ è la distanza massima su tutti i sottoinsiemi di Sn:

$$\|Q_1-Q_2\| = \max_{A\subset S_n} |Q_1(A) - Q_2(A)|$$

Si noti che questa è una misura molto sensibile della distanza tra le due densità. In primo luogo, si noti che $0 \leq \|Q_1-Q_2\| \leq 1$. Quindi, immaginiamo che U, sia la distribuzione uniforme su Sn, e  Q è una densità risultante da un particolare (generico) tipo di riordino. Se questo tipo di shuffle lascia una singola carta nel suo posto originale, si avrà:

$$\|Q -U\| = 1 - \frac{1}{n},$$
che è un valore alto dato che la distribuzione ordinata ha valore 1.
Per esempio dopo la prima mano nella procedura a 3 carte la distanza di probabilita' vale ancora $\frac{5}{6}$;

Si potrebbe desiderare di controllare quale sia la distanza iniziale tra la distribuzione iniziale Io, quella unitaria,   per tutti gli altri ordinamenti omogenei U, che corrisponde a:

  $$\| I - U \| = 1 - \frac1{n!}$$

Ebbene per esempio, anche dopo la prima mano nella procedura a 3 carte top card la distanza di probabilita' vale ancora $\frac{5}{6}$; ma successivamente la distribuzione di probabilità tende ad omogenizzarsi in tutte le configurazioni.
    Con questa definizione il nostro esempio indica che  $\|Q^k - U\| \to 0$  $k\to\infty$.

    In conclusione si può far vedere che la distanza per il k-esimo mescolamento tende a zero con k:

    $$ \|R^k - U\| \leq P(T>k) = 1 - \prod_{j=1}^{n-1}\left(1-\frac{j}{2^k}\right).$$

\begin{figure}[htbp]
\begin{center}
\includegraphics[scale=0.7]{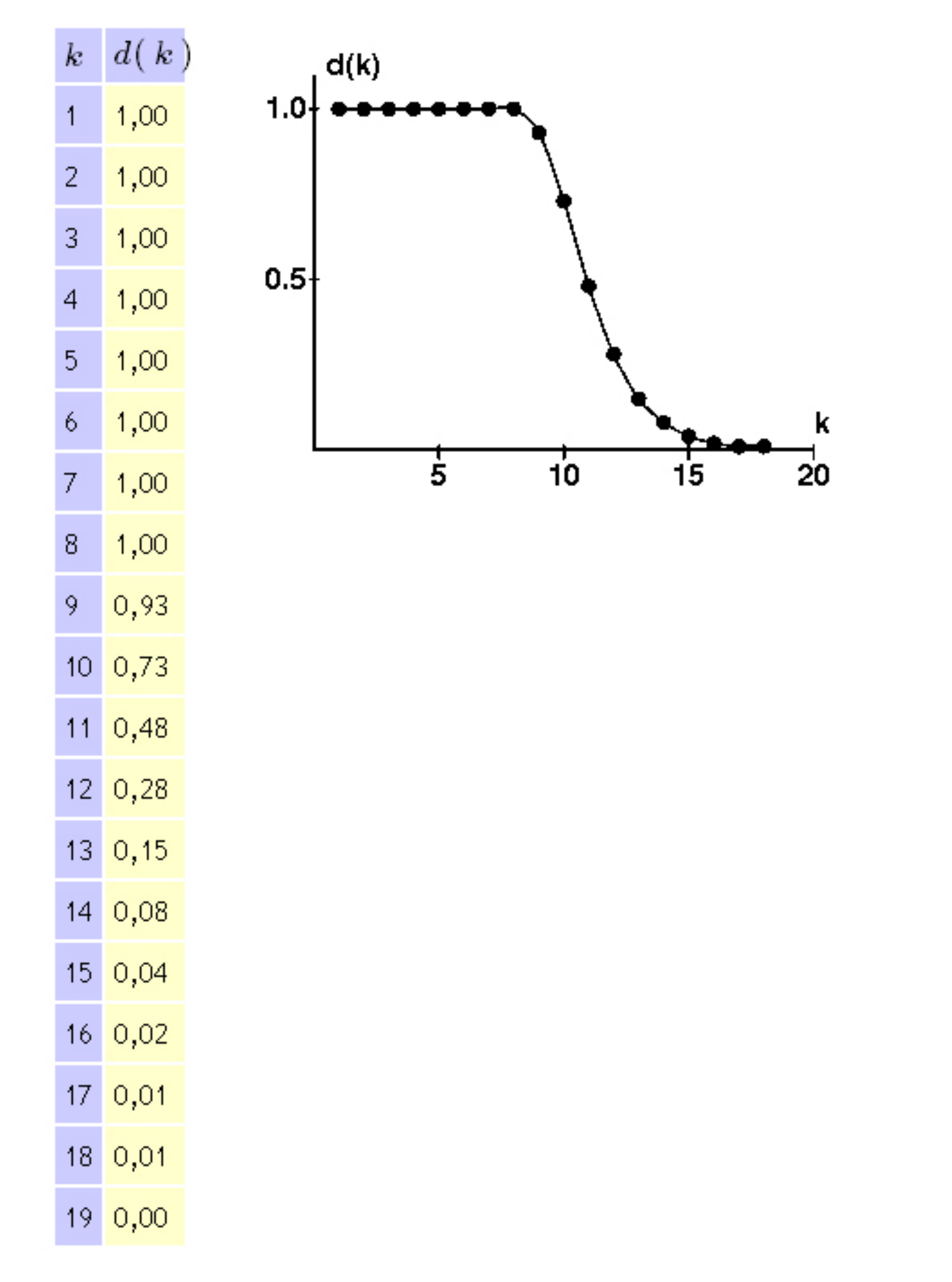}
\caption{La evoluzione della distanza per n=52 come in letteratura}
\label{12Riffle-GRAF.pdf}
\end{center}
\end{figure}

    Con questa procedure, consideriamo il mescolare un mazzo tipico di n = 52 carte  con mescolamenti all'americana (riffle shuffle).    Troviamo i limiti seguenti per gli andamenti sulla distanza tra il modulo del vettore $R^{k}$ e U, Che indicheremo con d(k) descritta in Fig:\ref{12Riffle-GRAF.pdf}

\begin{figure}[htbp]
\begin{center}
\includegraphics[scale=0.7]{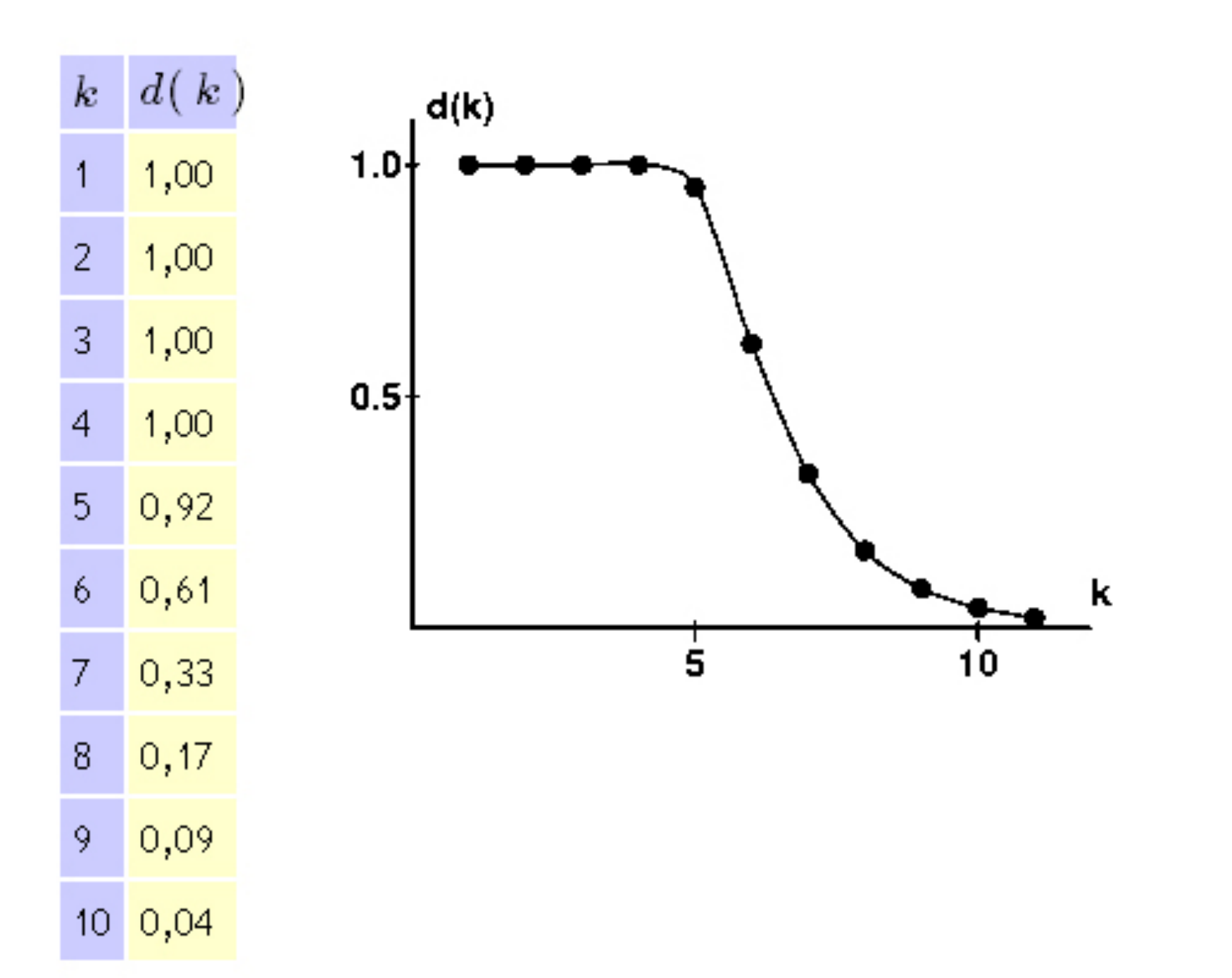}
\caption{}
\label{13Riffle_DIACONIS.pdf}
\end{center}
\end{figure}

Concludiamo questa presentazione ricordando che nei recenti lavori di Diaconis l'evoluzione prevista ha fornito una variazione della distanza piu rapida, come dalla sequenza seguente e dalla sua curva associata qui sotto: vedi Fig \ref{13Riffle_DIACONIS.pdf}.

  Si noti come rispetto alla precedente evoluzioni la $d(k)$ evolve più rapidamente (ovvero $d(k)$ critico più basso) rispetto a quella precedente.

\begin{figure}[htbp]
\begin{center}
\includegraphics[scale=0.7]{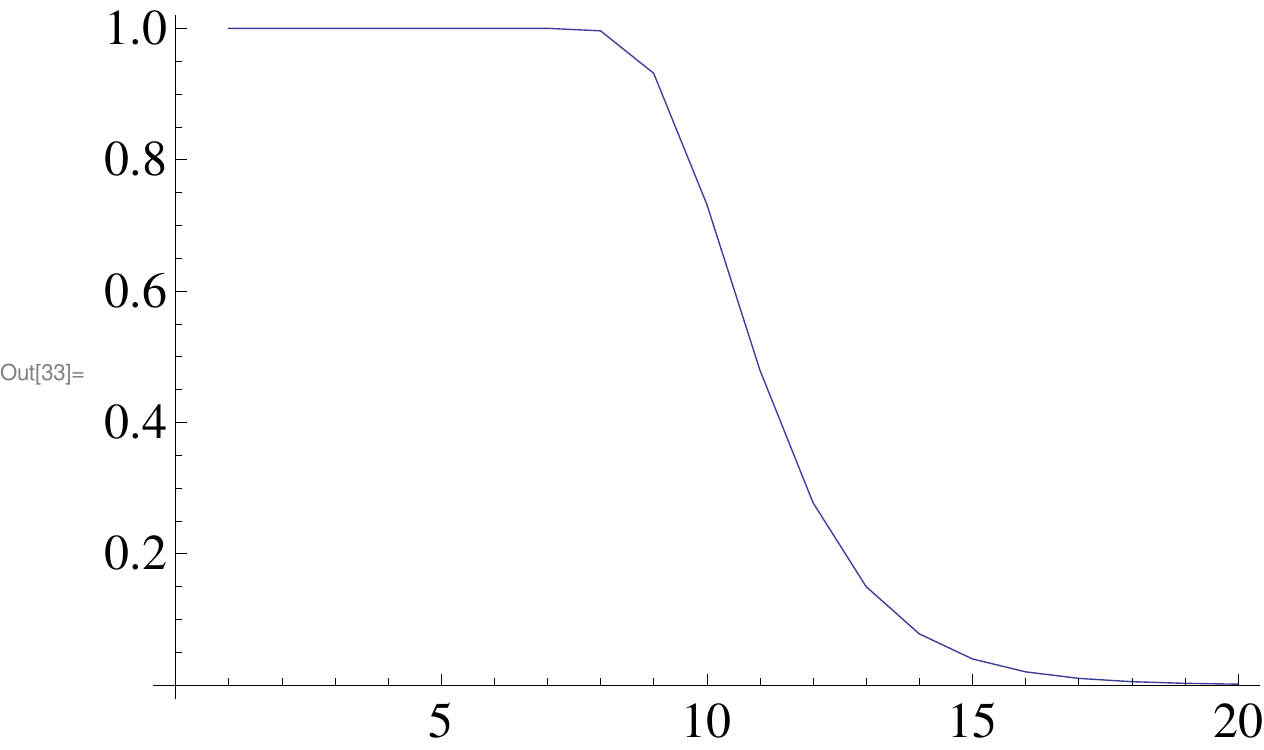}
\caption{La evoluzione della distanza da me calcolata per n=52}
\end{center}
\end{figure}

\begin{figure}[htbp]
\begin{center}
\includegraphics[scale=0.7]{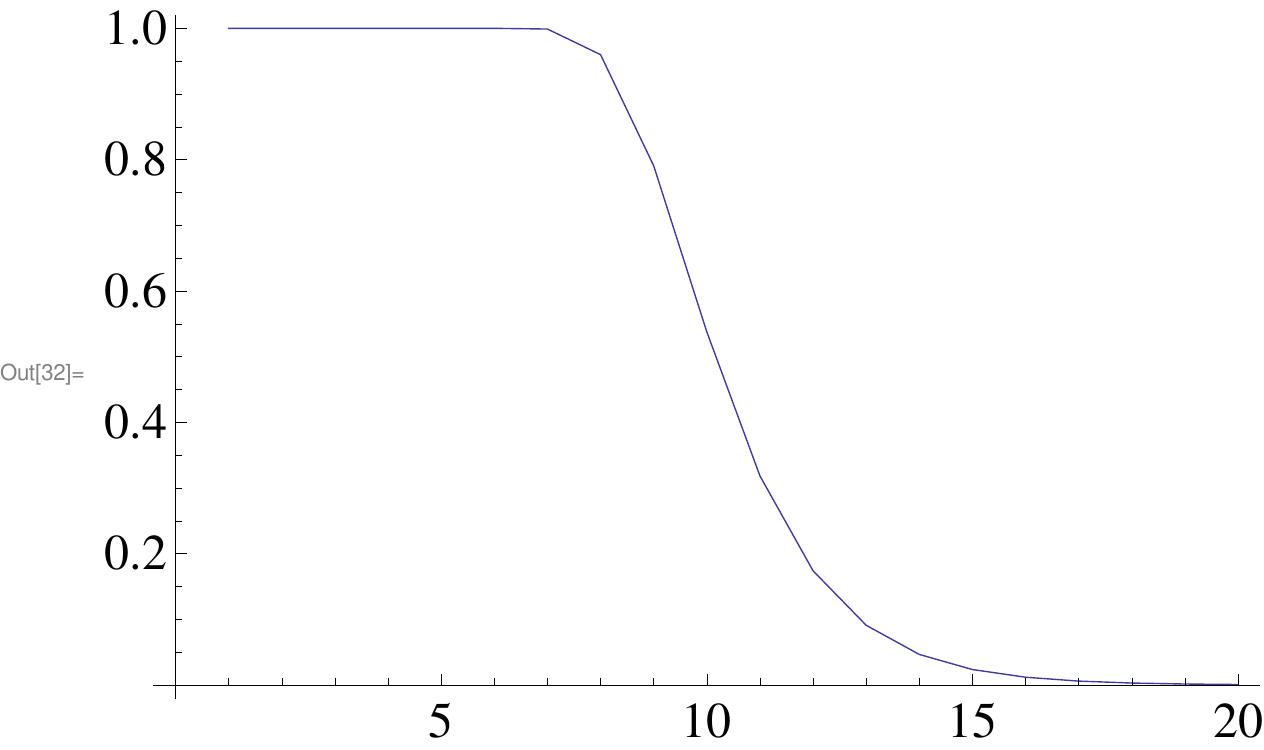}
\caption{La evoluzione della distanza da me  calcolata per n=40}
\end{center}
\end{figure}

\begin{figure}[htbp]
\begin{center}
\includegraphics[scale=0.7]{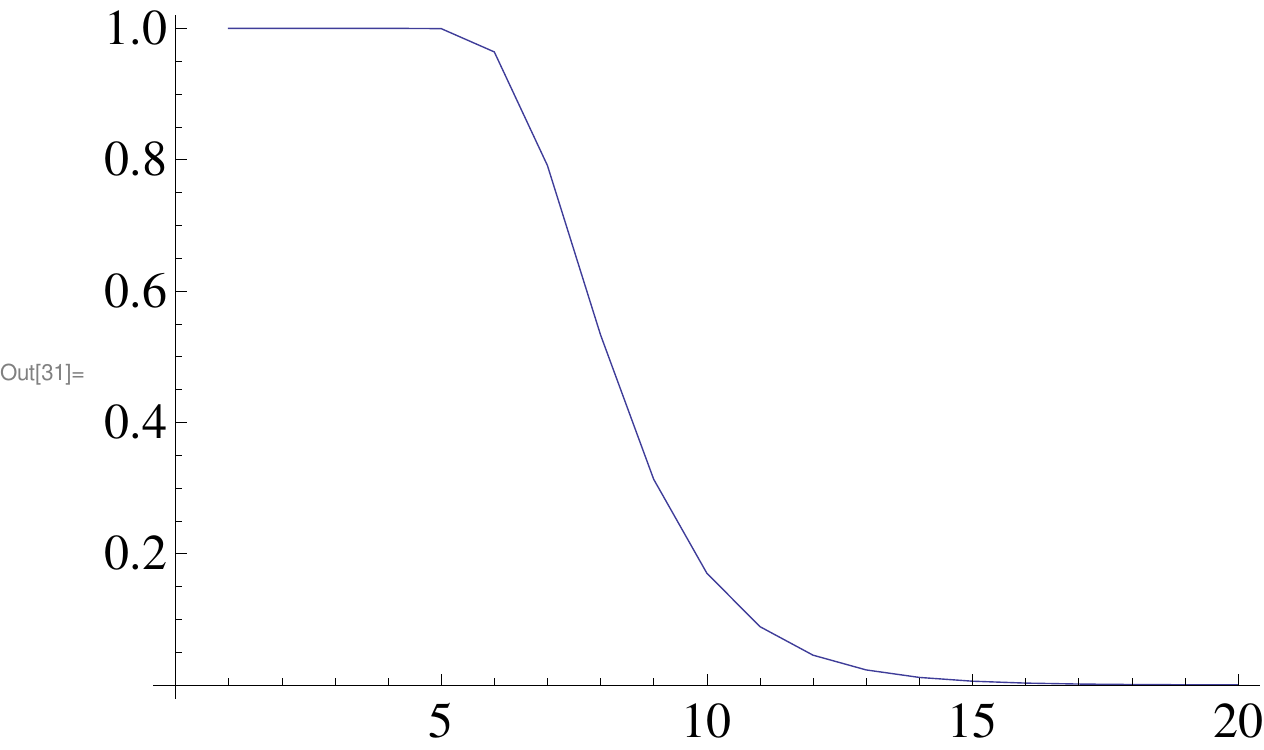}
\caption{La evoluzione della distanza da me calcolata per n=20}
\end{center}
\end{figure}

\begin{figure}[htbp]
\begin{center}
\includegraphics[scale=0.7]{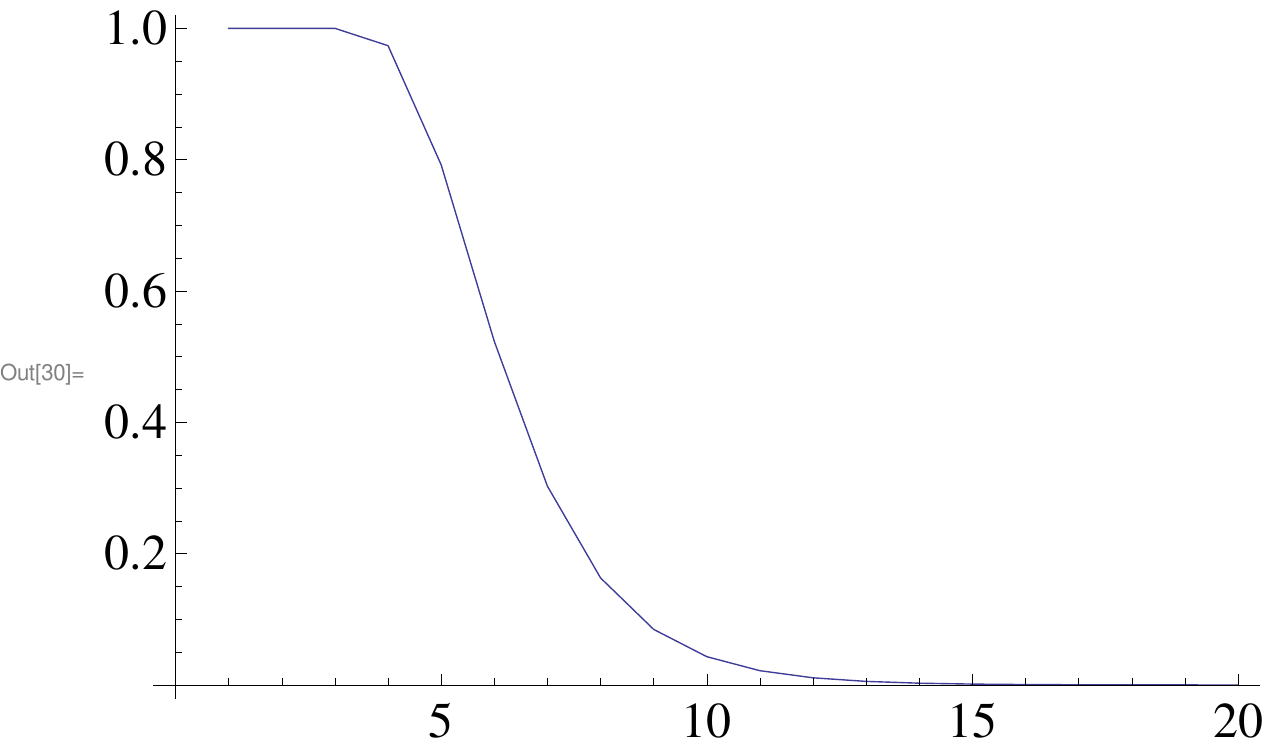}
\caption{La evoluzione della distanza da me calcolata per n=10}
\end{center}
\end{figure}

Con la presente  vediamo che d(k) è relativamente grande e costante per piccole k (in vero per top card $k < n \cdot (ln_{2} n$ )) ma essa diminuisce rapidamente  al crescere di k,
oltre al valore limite indicato (detto cut-off o di transizione). Vi è di conseguenza una regione di transizione critica   in cui d(k) passa da circa uno (o quasi,mantenendo un andamento costante) ad un decadimento simile al decadimento esponenziale.  Sembra ragionevole che esista un qualche valore critico  $k_{c}$, definito dal numero delle carte, che , allorquando la d(k) cominci a decadere, segnali il raggiungimento di un appropriato numero di mescolamenti : esattamente quanto  richiesto per garantire che il mazzo sia stato davvero ben mescolato. Utilizzando questo modello, risulta che k = 12 sarebbe sufficente per un buon mescolamento di 52 carte; si vede facilmente che $k_{c} \simeq ln_{2}n$. Andrebbe segnalato che si può arbitrariamente richiedere che la distanza $d(k)$ diventi un mezzo, un decimo oppure un millesimo ottenendo cosi tempi di mescolamento arbitrariamente diversi. Questo aspetto, direi microscopico, dell'analisi del decadimento lo si può evincere dallo studio della distanza $d(k)$ per 3 e 4 carte in scala semilog, analisi  che meglio  descrive la decrescita $d(k)$, di natura apparentemente esponenziale appena successiva al valore critico $k_{c}$(vedi anche capitoli successivi).

\begin{figure}[htbp]
\begin{center}
\includegraphics[scale=0.7]{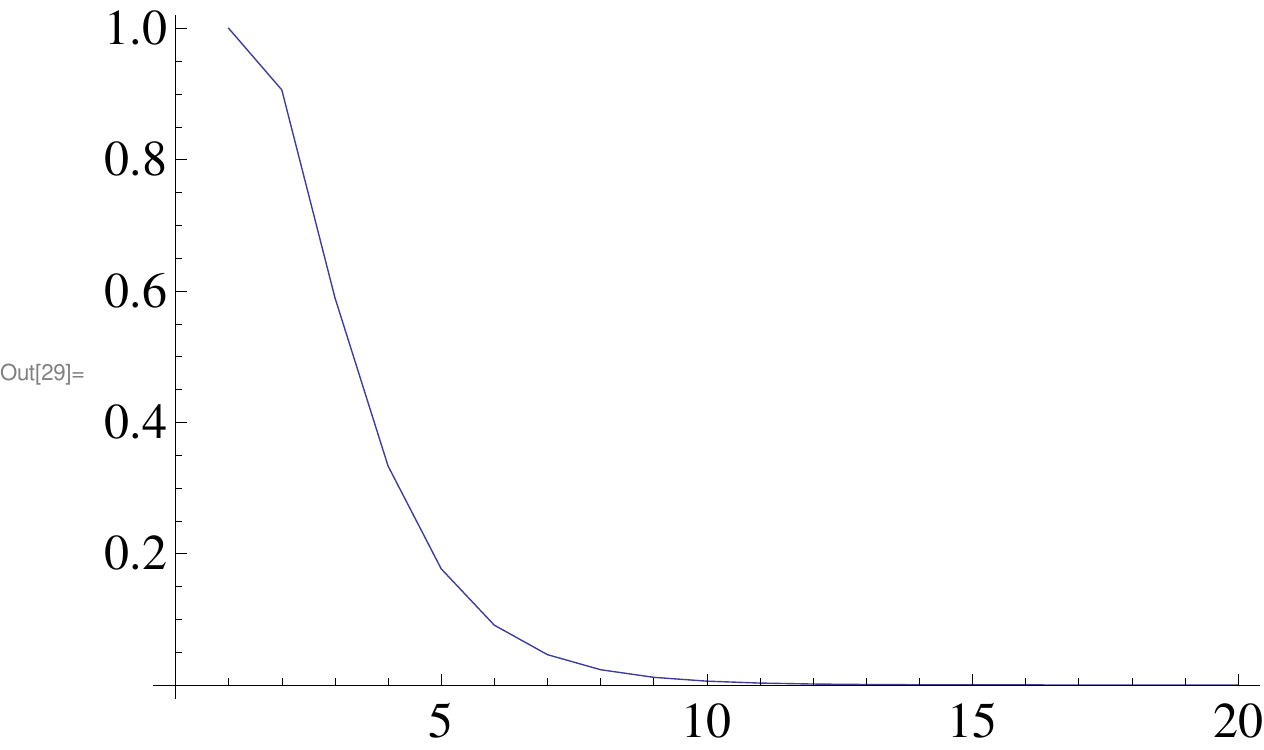}
\label{12fRiffle-GRAF.pdf}
\caption{La evoluzione della distanza da me calcolata per n=4}
\end{center}
\end{figure}

Nella mia tesi nel capitolo 3 ho provato analiticamente con 3 e poi 4 carte  dei risultati  discussi e descritti graficamente in dettaglio più avanti, relativi al decadimento d(k). In vero la riduzione d(k) della distanza  avviene come  nelle figure successive \ref{12fRiffle-GRAF.pdf}.

\begin{figure}[htbp]
\begin{center}
\includegraphics[scale=0.7]{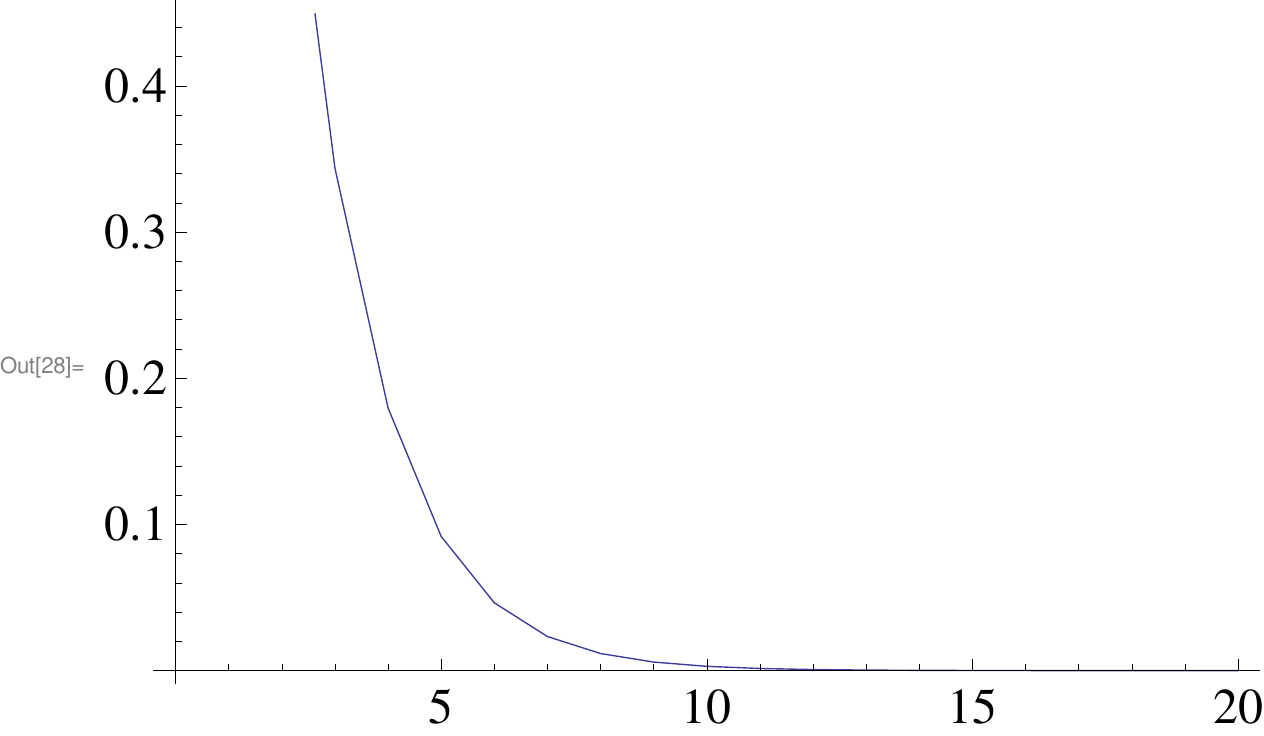}
\caption{La evoluzione della distanza da me calcolata per n=3}
\end{center}
\end{figure}

Queste evoluzioni (n $= 3,4$) sono oggetto dei capitoli successivi.

\chapter{Esempi numerici}
Consideriamo la distribuzione delle probabilità utilizzando il programma Mathematica (od un suo equipollente in C++).
Qui noi riproduciamo la stessa procedura del caso a 3 carte top card discussa sopra e il vettore delle probabilità per la prima mano (unità o identità),
il successivo vettore di un terzo o zero, e via via i successivi mescolamenti ed i vettori associati, che tendono al valor limite $\frac{1}{3!} = \frac{1}{6}$

Il Top Card Riffle  con  3 carte presenta le seguenti evoluzioni dei vettori di probabilità
ottenuti moltiplicando ogni volta il vettore unitario per le matrici $Q^{k}$ (o $R^{k}$) create dalle mix o shuffle top card successive:

\begin{figure}[htbp]
\begin{center}
\includegraphics[scale=1.1]{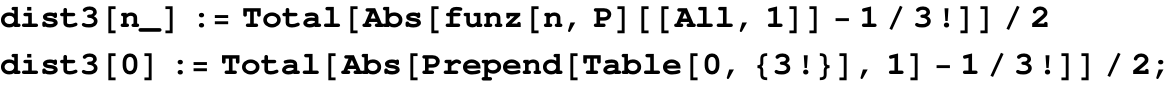}
\caption{procedura matematica per ottenere la distanza tra il vettore di probabilità e il vettore di uniformità in cui tutti e 6 gli elementi valgono $\frac{1}{3!} = \frac{1}{6}$}
\end{center}
\end{figure}

\begin{figure}[htbp]
\begin{center}
\includegraphics[scale=0.15]{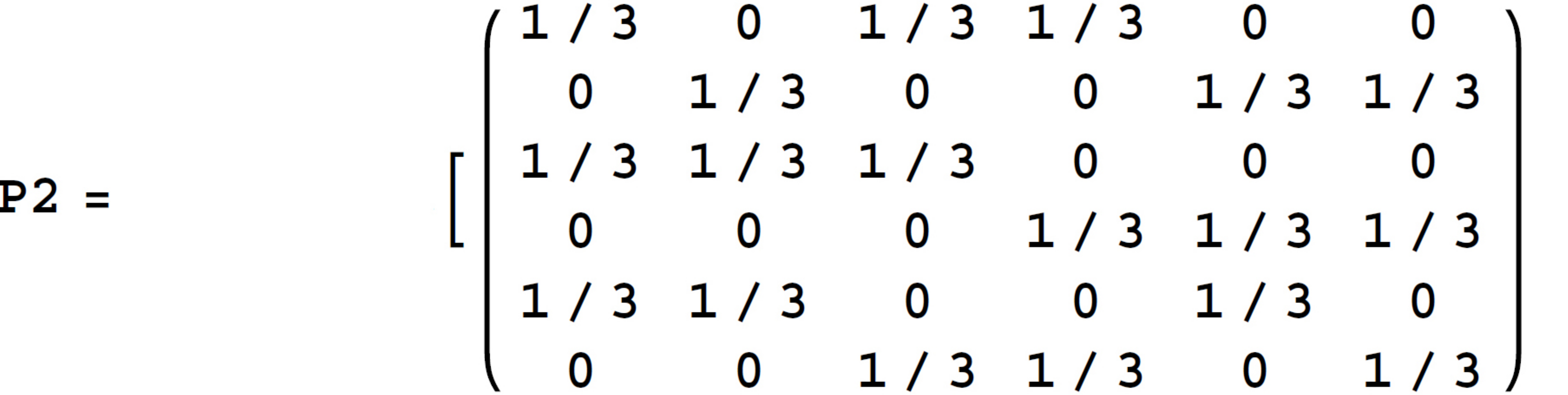}
\caption{ la matrice elementare (come descritto nel precedente capitolo) per le probabilità di mescolamento di sole 3 carte alla prima mano nella tecnica top card}
\end{center}
\end{figure}

\begin{figure}[htbp]
\begin{center}
\includegraphics[scale=1.6]{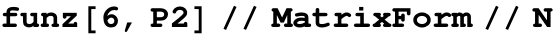}
\includegraphics[scale=0.6]{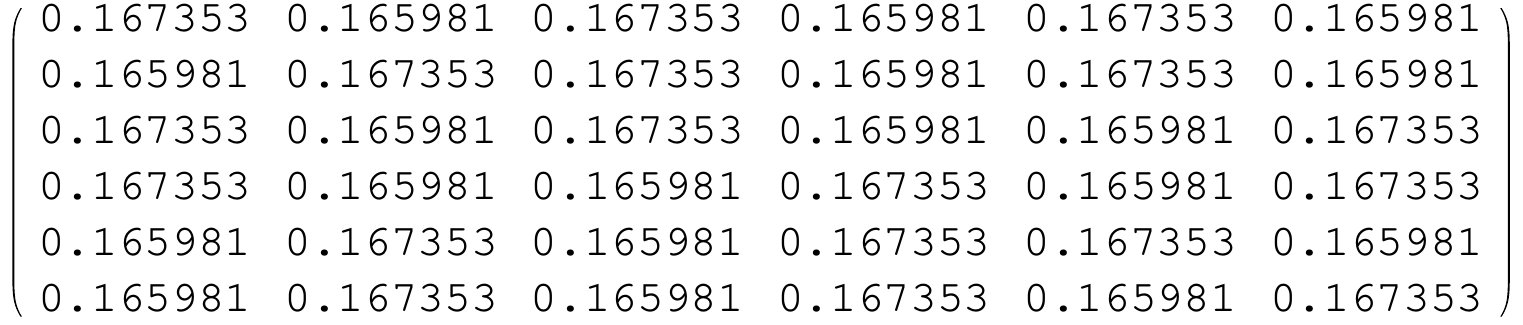}
\caption{Il comando di moltiplicazione delle matrici $P_{2}$ precedentemente calcolata   e la sua  evoluzione $Q^{k}$ matrice  corrispondente a 1,2,3..k mani con tecnica top card per 3 carte; qui siamo alla matrice  $Q^{6}$, ovvero l'operatore che deforma  il vettore di probabilità se mischiamo  ben 6 volte con tecnica top card; }
\end{center}
\end{figure}
 Questa procedura è diversa se considero il mescolamento riffle shuffle (all'americana): infatti esso mescola prima favorendo sempre il valore della prima configurazione come la più probabile; la matrice di mescolamento (la cui prima colonna nel caso iniziale coincide con il risultante vettore di probabilità) $Q^{1}$ che nell'elaborazione numerica  viene definita e calcolata analiticamente P diventa:
\begin{figure}[htbp]
\begin{center}
\includegraphics[scale=0.9]{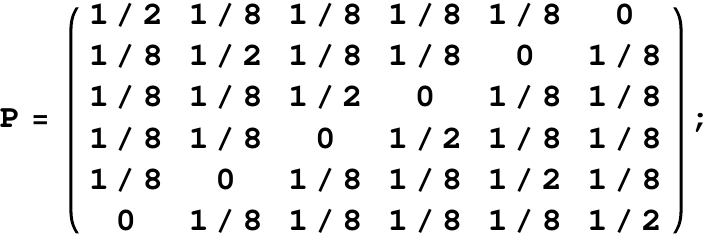}
\caption{La matrice che descrive una mescolata all'americana (riffle shuffle); sulla colonna sono sottintese le possibili permutazioni in ordine crescente: 123, 132,213..321; i valori di probabilità della matrice sono calcolati manualmente; la matrice viene definita quale matrice P}
\end{center}
\end{figure}

La successiva matrice di mescolamento, ovvero $Q^{2}= P^{2}$
mantiene ancora il termine diagonale alto più simile al valore iniziale.
Questa matrice moltiplicata sul vettore unitario k volte produce i seguenti vettori.
Essi anche alla decima mano sono ancora leggermente
(quarta cifra) dissimili dal valore limite di un sesto.
Come si osserva nella sucessiva operazione di prodotto di Matrice $Q^{3}$

Adesso analizziamo il caso a 3 carte gia' discusso nel precedente capitolo, la cui matrice definiamo definiamo P2:

\begin{figure}[htbp]
\begin{center}
\includegraphics[scale=0.5]{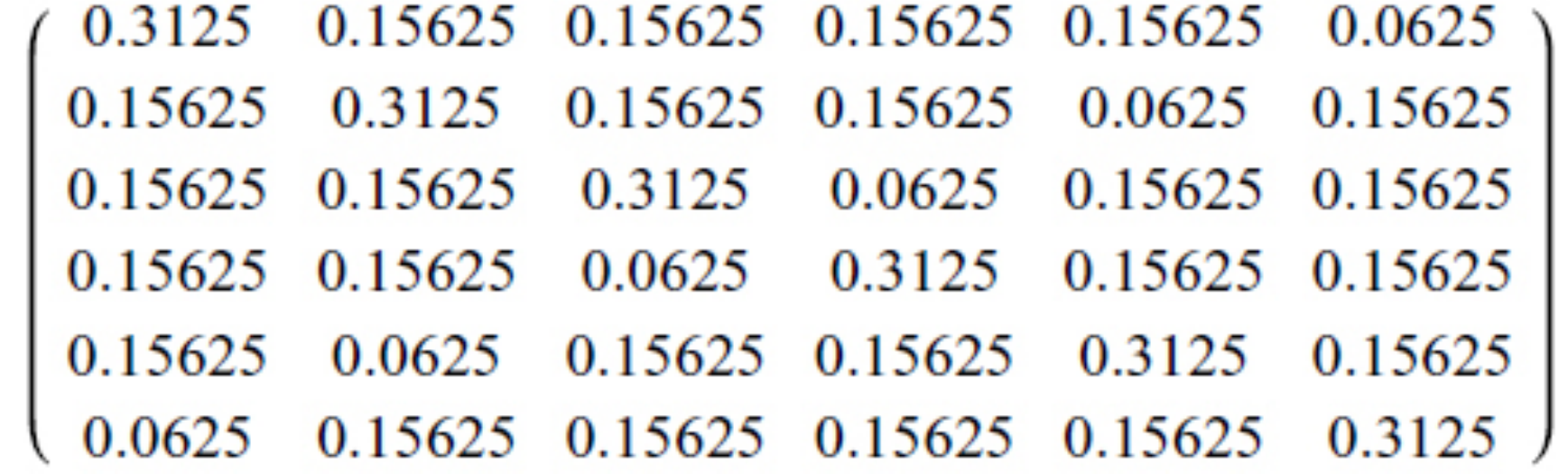}
\caption{Matrice di probabilità per la seconda mano del riffle shuffle per  3 carte}
\includegraphics[scale=0.6]{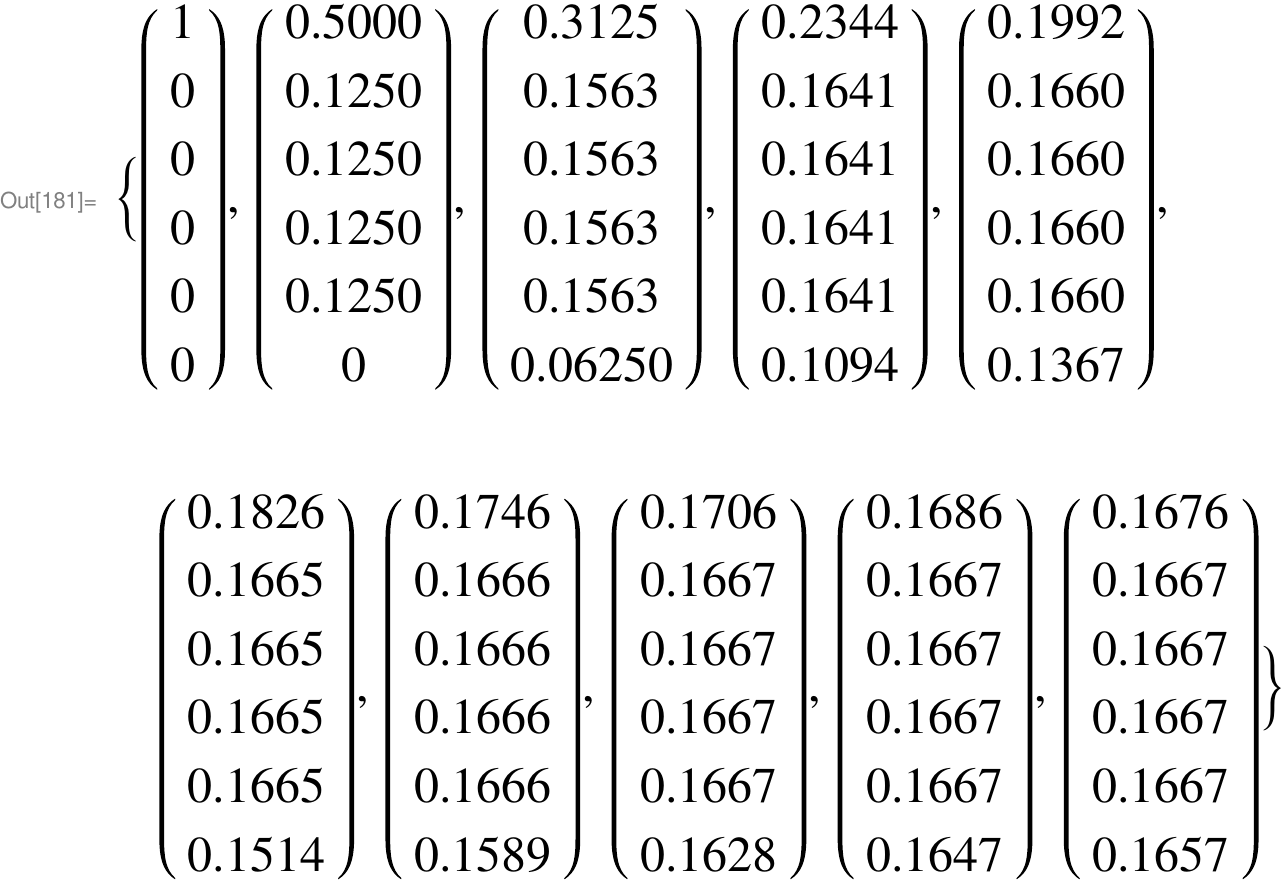}
\caption{Evoluzione della prima colonna, il vettore derivato dalla matrice $P^{k}$ for k=0,1,2,3.. per le prime dieci mani riffle shuffle. Notare che a differenza del top card shuffle la probabilità di ritrovare la prima carta in alto nei  k mescolamenti rimane più probabbile (ovvero ha il massimo del valore).}
\end{center}
\end{figure}

Le configurazioni di evoluzione delle probabilità $P^{6}$ qui ottenute per sei mani riffle su tre carte vengono derivate dal comando:

Tali risultati vengono ora estesi al caso a 4 carte.

\section{Mischiando 3 o 4 carte in top card shuffle}
Consideriamo ora il caso delle 4 carte richiamando per un confronto brevemente i risultati del mescolamento all'americana (riffle shuffle) a 3 carte ; le configurazioni per 4 carte sono $4! = 24$ e possono quindi essere descritte da vettori lunghi 24 elementi e matrici assai ingombranti di $24^{2}= 576$ elementi. Nella seguente si considera la distribuzione di probabilità nel caso di mescolamento top card sia del caso riffle shuffle.
Si possono facilmente trovare tali componenti nel caso top card.Questa matrice al quadrato permette di calcolare le probabilità finali (e sottraendole dall'unitaria ottenerne la distanza). Come gia' indicato la prima colonna della matrice $Q^{k}$ indica la evoluzione del vettore di probabilità nelle 24 configurazioni.
Ricordiamo che la distanza dalla omogeneità di cui vedremo l'andamento in funzione del numero di mani si deriva dalla equazione:

$$d= \|Q -U\| = \| Q^{k} - I_{o}\frac{1}{n!}\|.$$

Questa distanza come anticipato tende a zero per $k\longmapsto \infty$.
Nel presente paragrafo e nelle conclusioni vedremo in dettaglio in che modo in scala semilog questo avvenga.

La evoluzione della matrice  $Q^{k}$ di mix $4!$ da noi elaborate analiticamente
come dalla figure precedenti per 3 carte,  parte da una matrice ordinata nella prima mano, e nella tecnica top card si ottiene:

\begin{figure}[htbp]
\begin{center}
\includegraphics[scale=0.6]{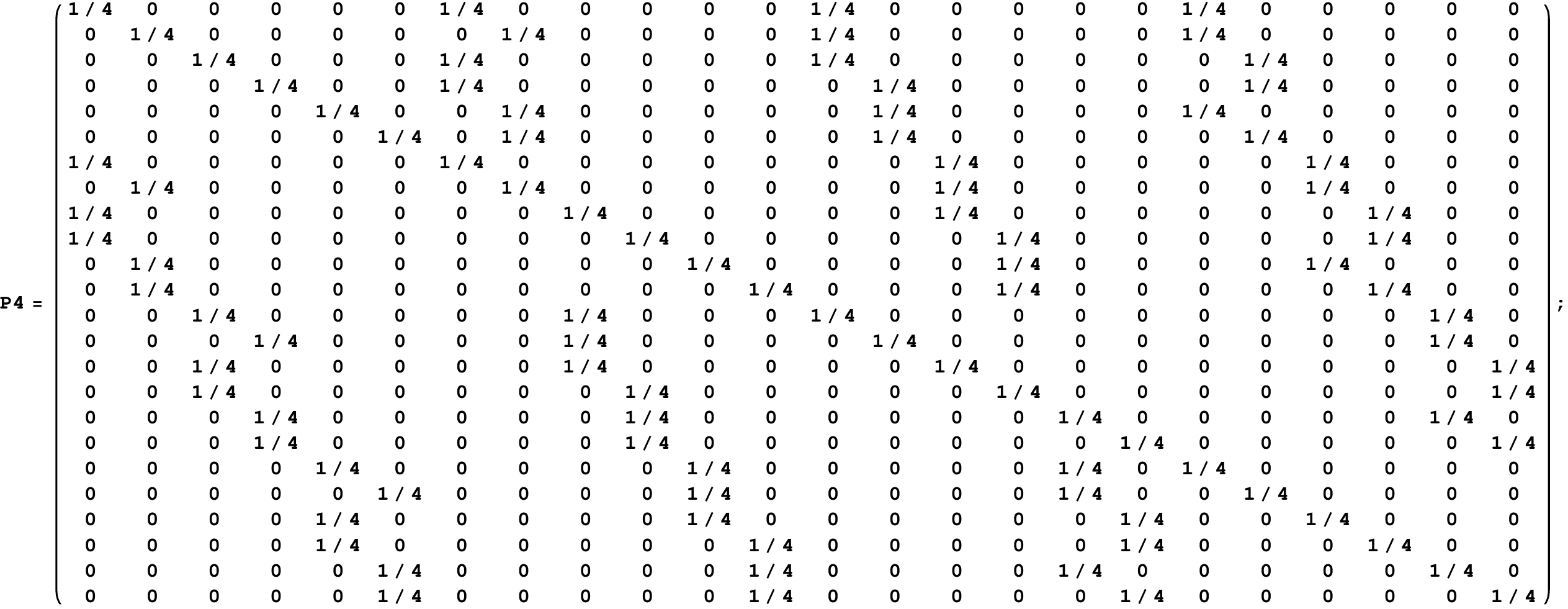}
\caption{Matrice  (per mescolamento top card)  di probabilità calcolata analiticamente; sulle colonne sono immaginati in ciascuna delle 24 configurazioni le permutazioni: 1234, 1243, 1324,1342...in crescendo fino a 4321;
i valori indicati sulla matrice sono le probabilità di ottenere un'altra configurazione dalla prima mano di mescolamento; questa matrice viene nominata P4 nel programma; il primo vettore permette di stimare  d pari a $d= \frac{20}{24}= \frac{5}{6}$}
\end{center}
\end{figure}

e successivamente alla seconda  mano, ottenuta quadrando la precedente matrice $Q$,

\begin{figure}[htbp]
\begin{center}
\includegraphics[scale=0.4]{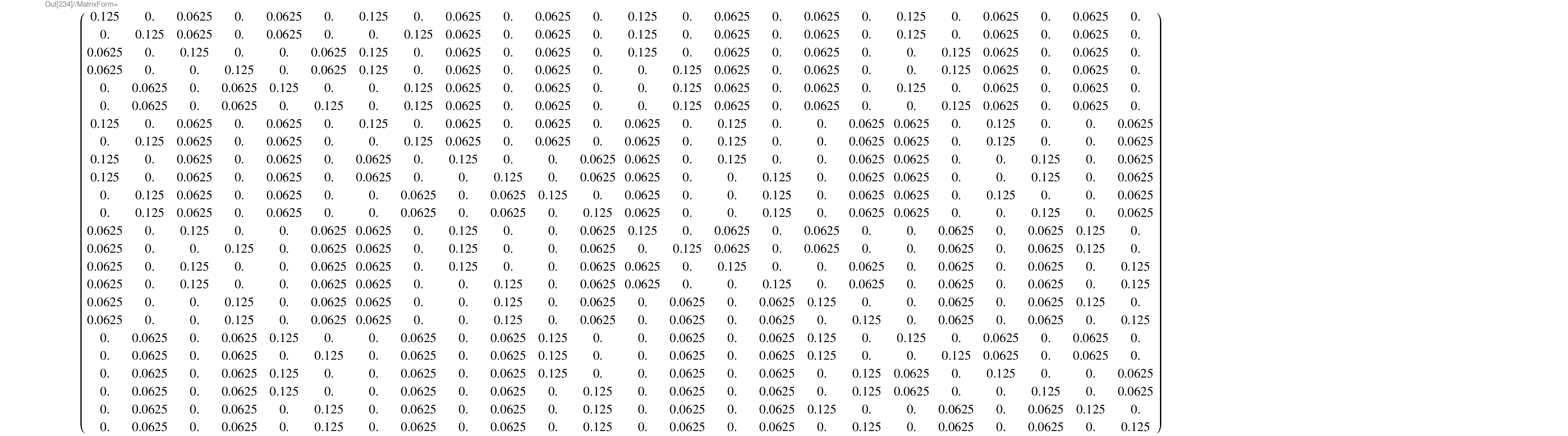}
\caption{}
\end{center}
\end{figure}

Senza riportare tutte le matrici ottenute , ma indicando solo la prima colonna, sintetizziamo meglio l'evoluzione del mescolamento (top card):

\begin{figure}[htbp]
\begin{center}
\includegraphics[scale=0.45]{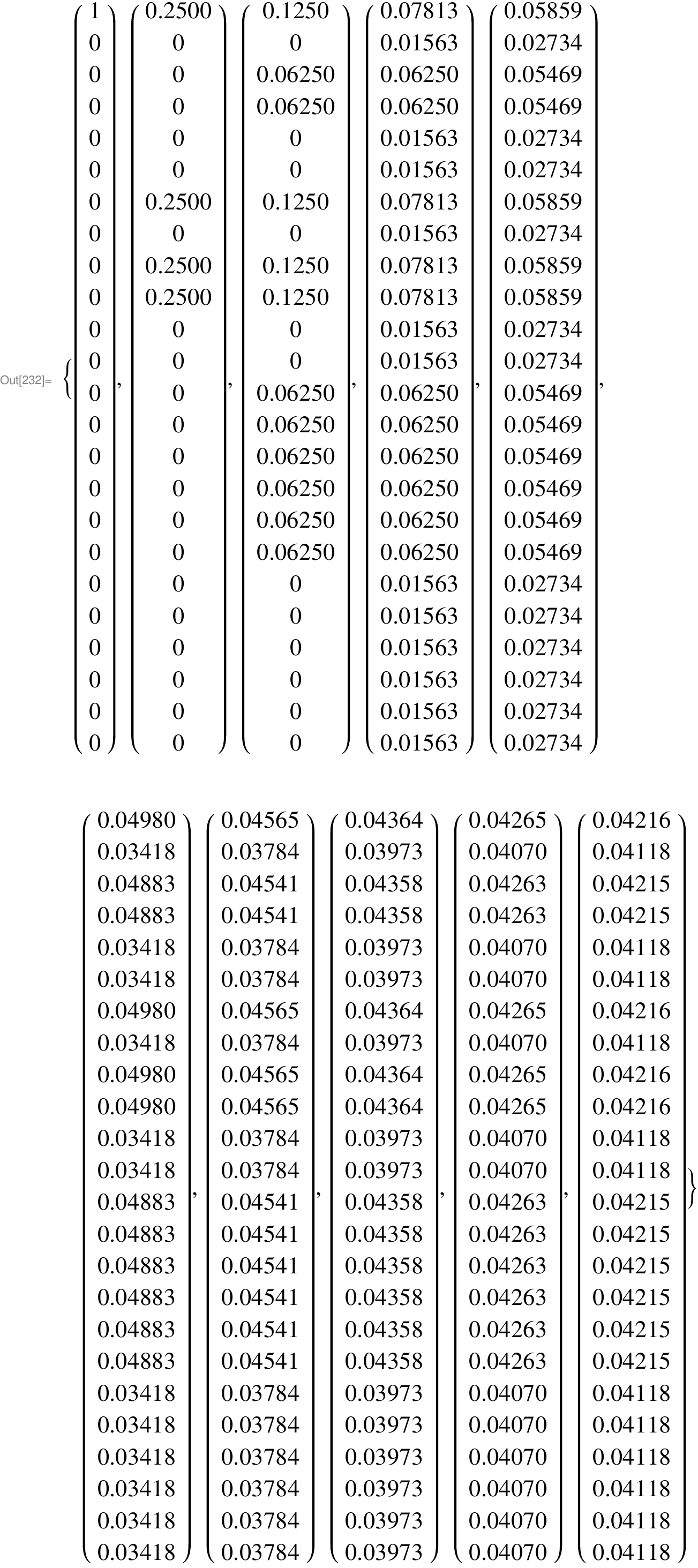}
\caption{Mescolamento delle probabilità per un top card di 4 carte fino alla decima mano. Questa evoluzione
indica come le varie configurazioni tendano al valore  finale $\frac{1}{4!} = 0.416\bar{6}$.}
\end{center}
\end{figure}

\begin{figure}[htbp]
\begin{center}
\includegraphics[scale=1.5]{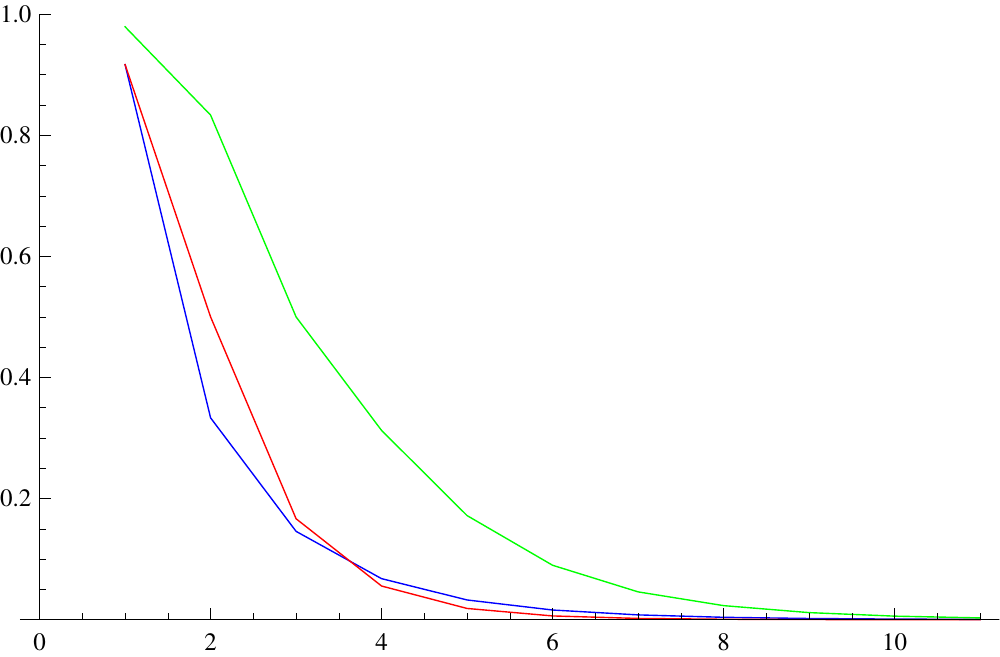}
\caption{La evoluzione della distanza $d(k)$ con il numero di mescolamenti: il blue indica il Top card che supera la curva rossa (riffle shuffle) mentre la curva verde indica la top card a 4 carte; si noti il sorpasso del metodo top card rispetto riffle shuffle a circa la quarta mano. Vedremo in scala semi-log in dettaglio questa evoluzione.}
\end{center}
\end{figure}
  Si noti come le probabilità restino fortemente variabili all'inizio  ma vadano a valori limiti pari a $\frac{1}{4!}= \frac{1}{24}\simeq 0.041666$.
  Si possono rappresentare in grafico lineare (in funzione di k) le distanze $d(k)$ definite nel capitolo precedente. In esso sono presenti le due curve piu interne (blue e rossa) per il top card e riffle shuffle
  a tre carte ed il caso top card a 4 carte (verde) più esterno.
  Si osserva un curioso sorpasso nel caso a tre carte ed una evoluzione di decadimento esponenziale per tutte e tre.  Le andiamo a confrontare meglio in scala semilogaritmica nel capitolo successivo.

 \chapter{Conclusione: Mischiando 4 carte riffle shuffle}

Il problema della natura disordinata del mazzo di carte s'associa al concetto d'approssimazione della configurazione ottenuta nel caso di massima uniformità, caso che abbiamo analizzato nei capitoli precedenti. In questo lavoro si sono considerate configurazioni a 3 carte (già note e presenti in letteratura)
  e configurazioni a 4 carte analizzate analiticamente e non presenti  allo stato attuale delle nostre conoscenze, in letteratura.  Tali analisi indicano, come ovvio, che il mix top card è più lento del riffle shuffle nel raggiungere l'omogeneità,   ma sorprendentemente, nelle fasi finali dello shuffle a 3 carte, avviene un sorpasso: la distanza dalla omogeneità sembra  raggiunta meglio dalla top card shuffle.
    Viceversa per il caso delle 4 carte la convergenza alla omogeneità avviene più lentamente   che nel caso a 3 carte; infatti in generale ci attendiamo che la transizione al disordine debba avvenire per $n \ln_{2}n$ per il top card e (seguendo Diaconis) attorno $\frac{3}{2}\ln_{2}n$  per il riffle shuffle, come espresso nella rinomata pubblicazione  la cui prima pagina introduttiva viene qui inserita. Inoltre  lo studio delle matrici $Q^{k}$   di mescolamento per 4 carte presentano  interessanti strutture di simmetria e leggi di decadimento descritte chiaramente nelle figure seguenti in formato semi-log.

\begin{figure}[htbp]
\begin{center}
\includegraphics[scale=0.32]{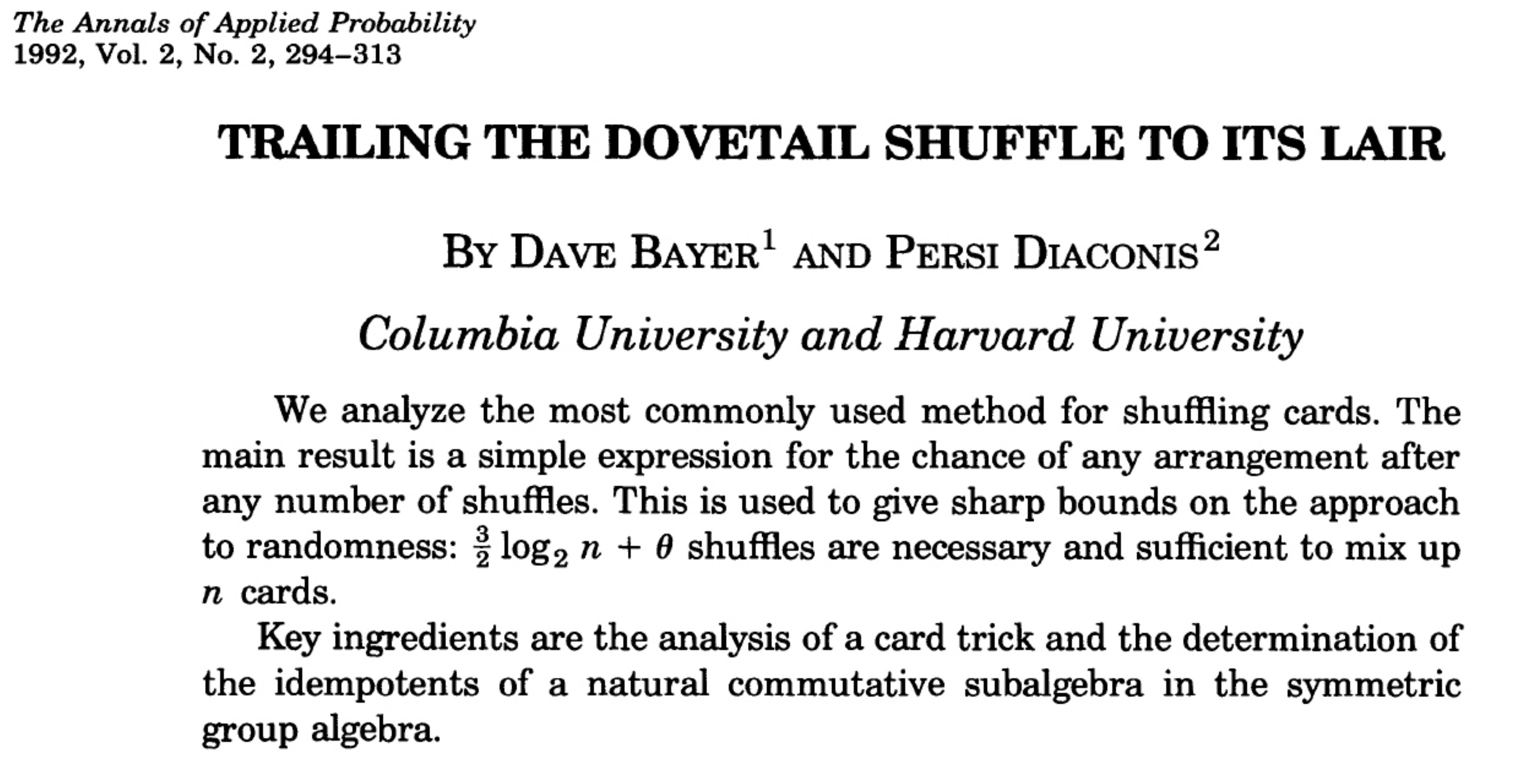}
\caption{Abstract dell'articolo 1992 di Bayer e Diaconis sulla transizione al disordine per il ruffle shuffle;
in esso si nota la distanza critica per la transizione all'omogeneità, $d_{c}= \frac{3}{2} ln_{2}n$.}
\end{center}
\end{figure}

Da tali matrici posso quindi effettivamente determinare  la evoluzione del vettore di distanza di probabilita',
i cui valori tendono effettivamente a $\frac{1}{4!}$ in circa $4\cdot log 4$ passaggi.
Ma tale avanzamento procede in scala semilog abbastanza a lungo. In pratica fin tanto che le probabilità stanno cambiando fortemente, la distanza  è grande, mentre se l'approssimazione alla omogeneità si avvicina, allora il decadimento e' esponenziate, vicino alla potenza di 2. Nella figura seguente la curva verde con i pallini indica il top card shuffle di 4 carte mentre le due curve interne rosso e blu indicano  rispettivamente la top card e la riffle shuffle mix per 3 carte. Ricapitoliamo  quanto ottenuto per il top card a 3 e 4 carte
e poi consideriamo il caso top card e riflle per 3 e 4 carte.
Si noti il sorpasso nella omogeneita' della top card rispetto alla riffle per le 3 carte.Si noti anche che per il riffle la evoluzione di probabilita' ottenuta, anche alla decima mano, mantiene una maggior probabilità trovarsi nella stessa configurazione di partenza.

\begin{figure}[htbp]
\begin{center}
\includegraphics[scale=0.65]{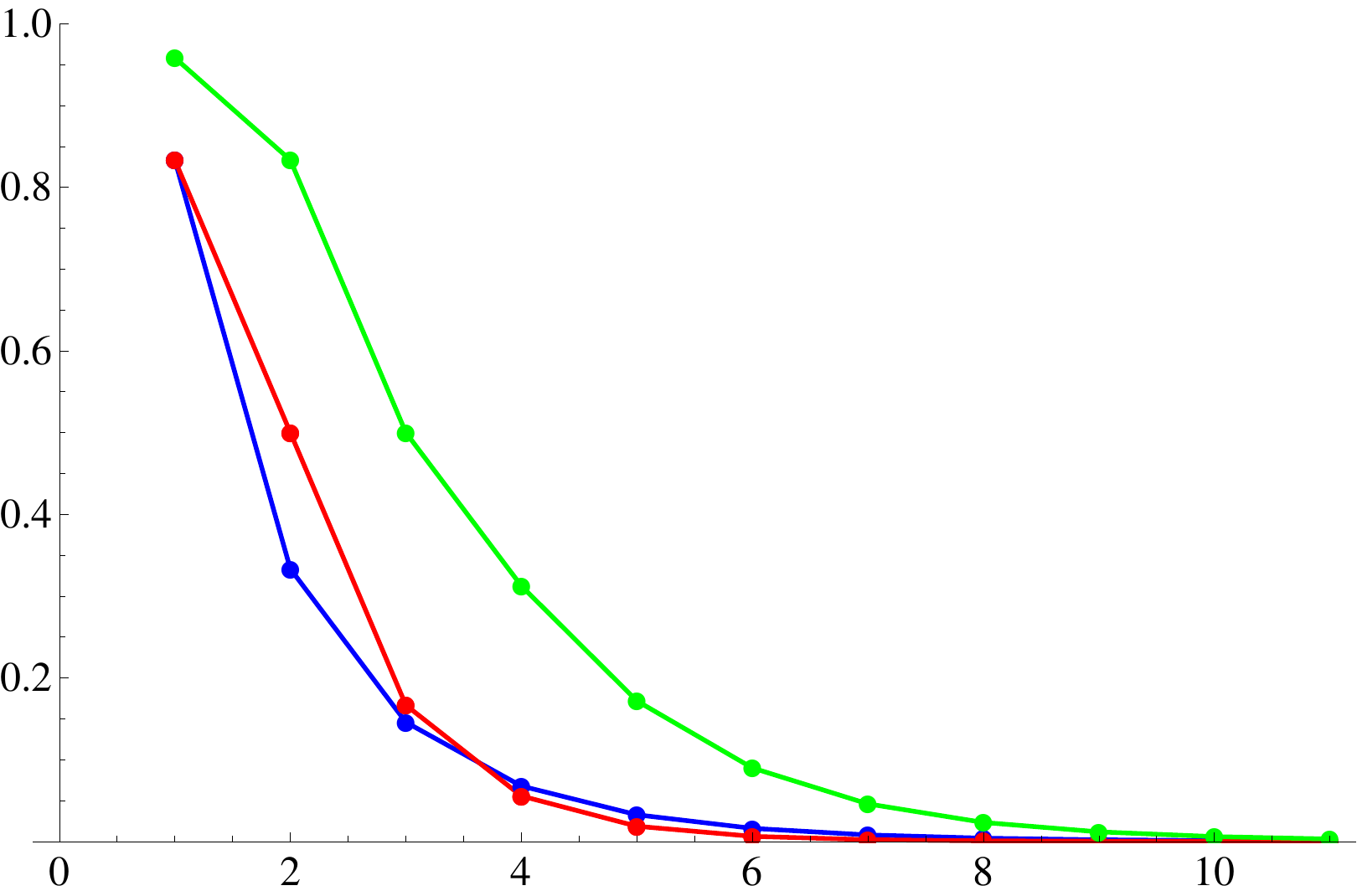}
\caption{Primo grafico che considera il decadimento della distanza dall'omogeneità $d(k)$ in funzione lineare del numero di mescolamenti k. La transizione nel caso 3 e 4 dal 1 allo 0 non si puo evidenziare; per 52 carte il processo di transizione avviene più chiaramente. Nelle successive analisi vedremo di studiare in dettaglio l'evoluzione $d(k)$ dall'ordine $1$ al disordine completo, ovvero $d(k)= 0$.}
\end{center}
\end{figure}

\section{Distanza dall'uniformità,\\
 riffle shuffle per 3 e 4 carte}

\begin{figure}[htbp]
\begin{center}
\includegraphics[scale=0.6]{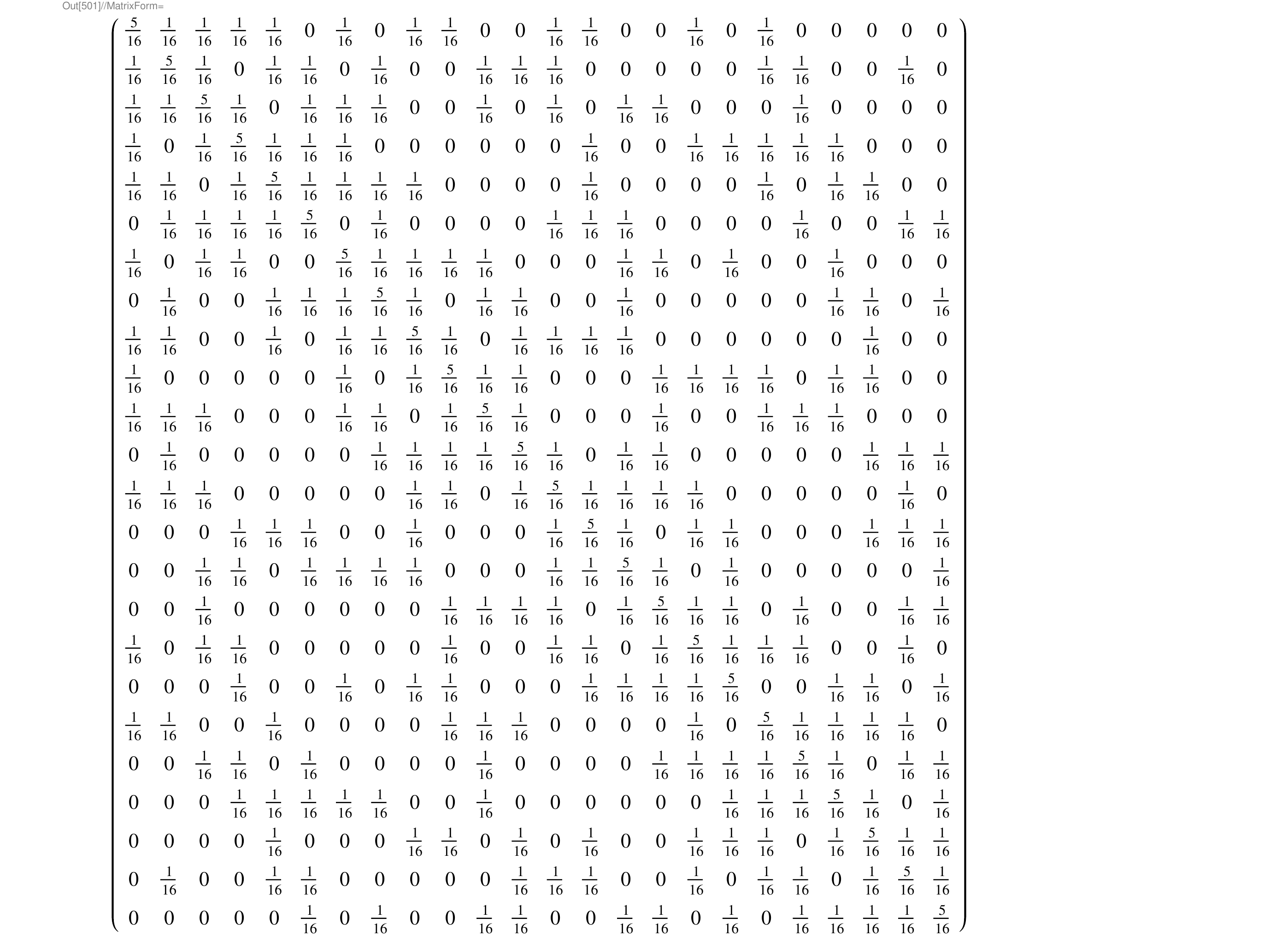}
\caption{La evoluzione della matrice a 4 carte per mescolamento shuffle alla prima mano: si noti come i valori in diagonale  siano maggiori delle altre componenti della matrice; questo indica una maggior probabilità di ritrovare la configurazione iniziale; la distanza d dalla uniformità per la prima mano come si può calcolare facilmente è pari a $d = \frac{1}{2}$}\label{24RIFFLE}
\end{center}
\end{figure}

Come si puo notare la matrice  riffle shuffle è diversa dal caso
top card precedente top card; vedi Fig. \ref{24RIFFLE}.

\begin{figure}[htbp]
\begin{center}
\includegraphics[scale=0.35]{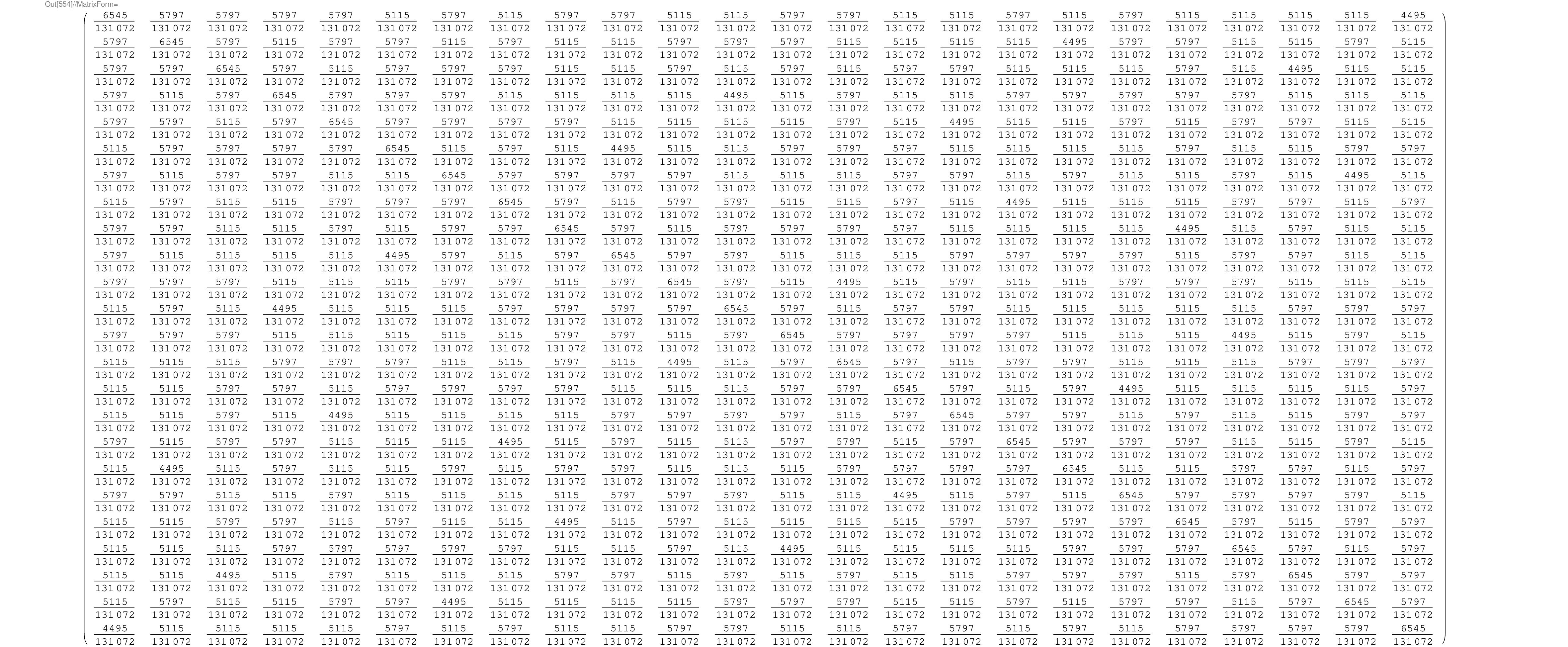}
\caption{Valore della matrice alla quinta mano in shuffle riffle   per 4 carte}\label{24RIFFLE5}
\end{center}
\end{figure}

Alla quinta mano abbiamo una matrice molto più complessa: vedi Fig.\ref{24RIFFLE5}.

In generale per i vettori per ben dieci mani  otteniamo la sequenza descritta in Fig. \ref{24RIFFLE10}:

\begin{figure}[htbp]
\begin{center}
\includegraphics[scale=0.5]{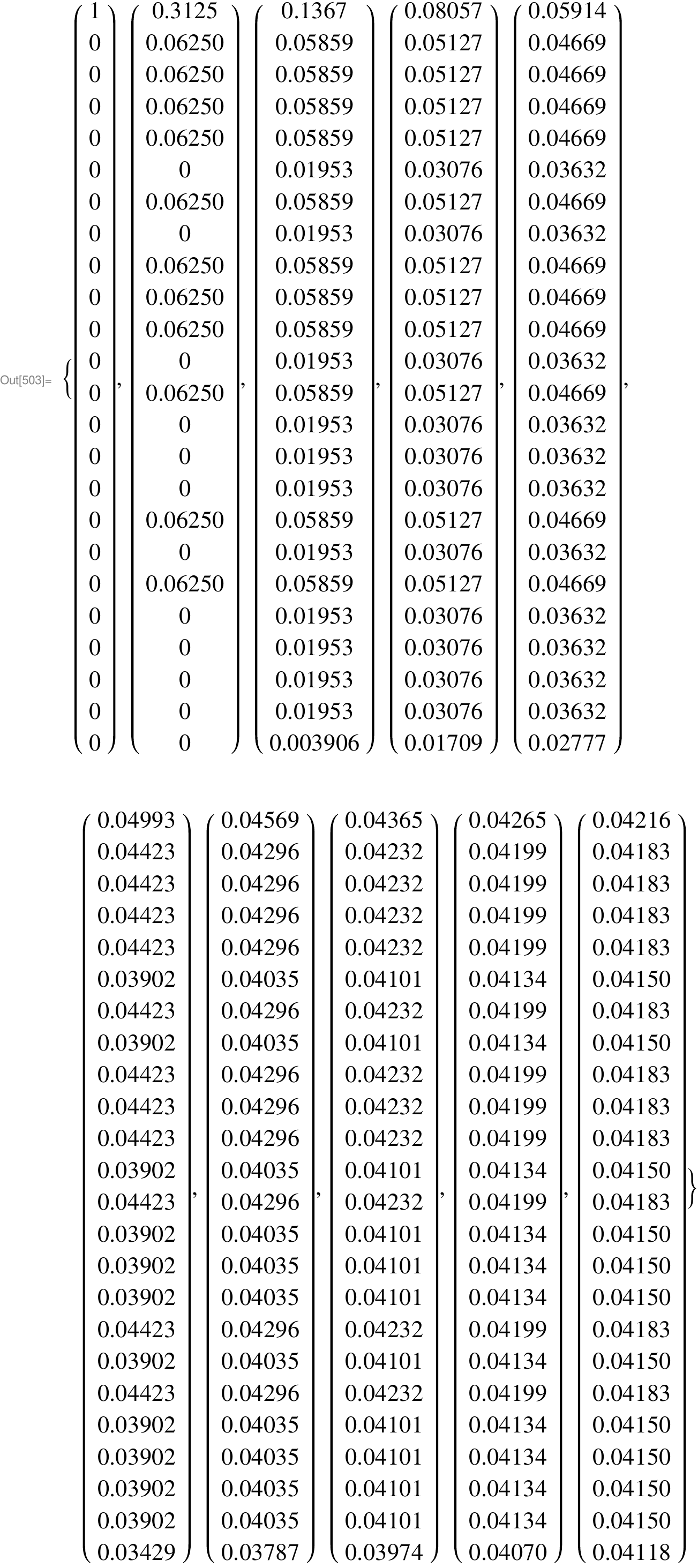}
\caption{La evoluzione dei vettori di probabilità per 10 mani successive per 4 carte in riffle shuffle; si noti come si proceda lentamente alla omogeneità ovvero a $\frac{1}{4!}$, come descritto graficamente nelle figure successivamente.}\label{24RIFFLE10}
\end{center}
\end{figure}

\begin{figure}[htbp]
\begin{center}
\includegraphics[scale=0.6]{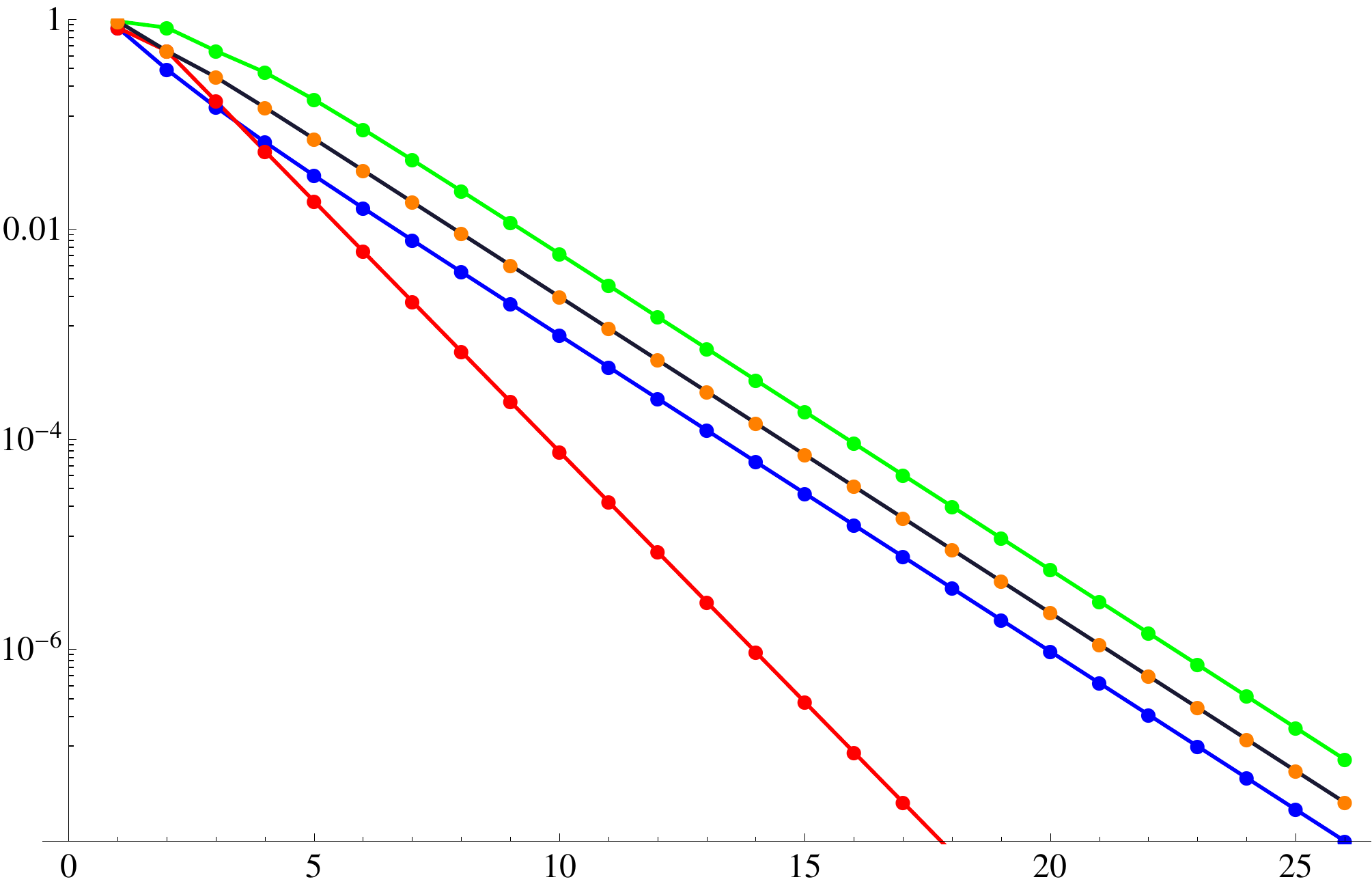}
\caption{La sequenza delle distanze in scala semilog dello shuffle a 3 carte top card (in rosso) e a riffle (in blue); accanto anche le evoluzioni rispettive a 4 carte sia top card (verde) che  riffle shuffle (nera); }
\end{center}
\end{figure}
\begin{figure}[htbp]
\begin{center}
\includegraphics[scale=0.6]{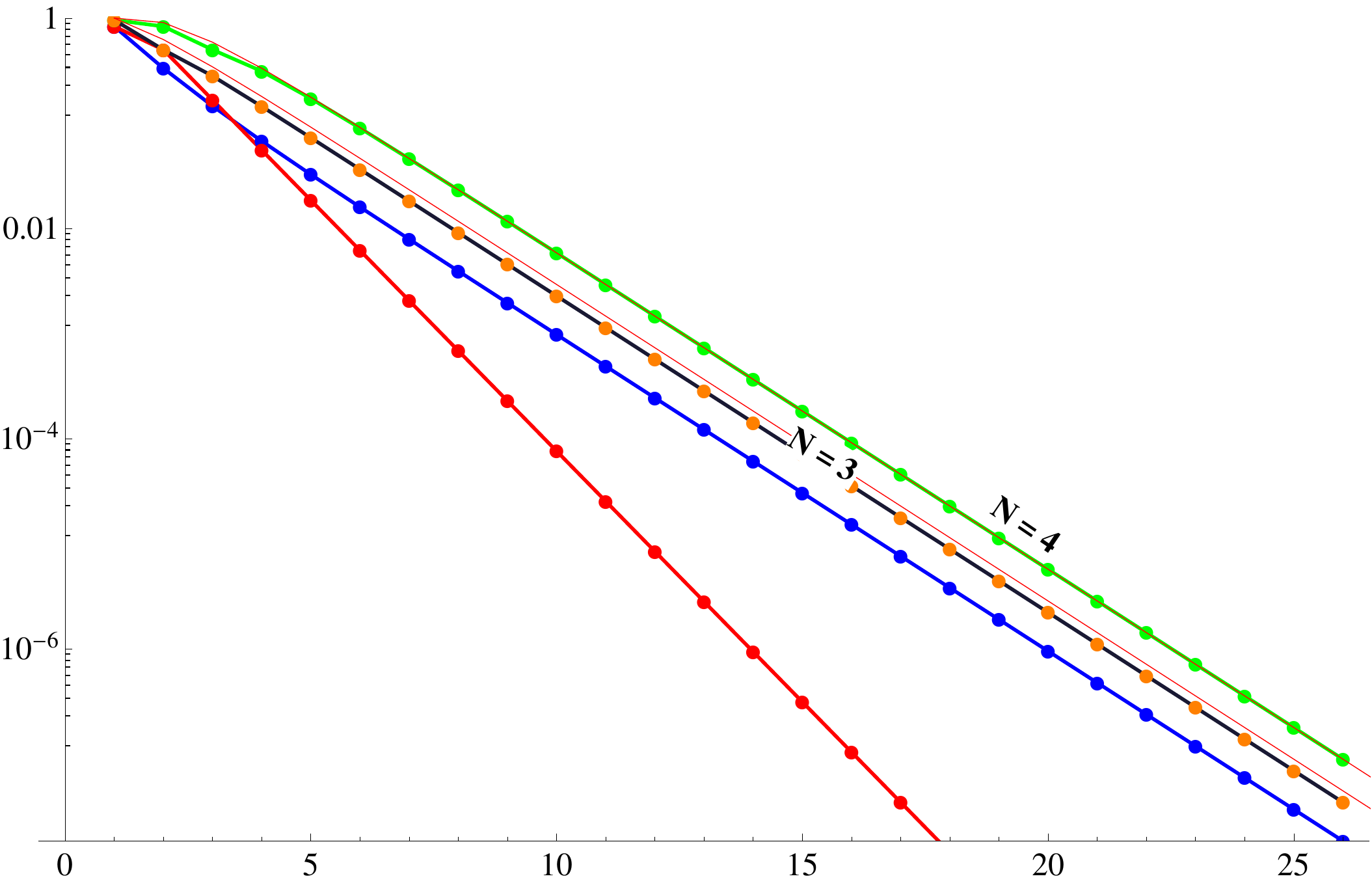}
\caption{La sequenza delle distanze in scala semilog dello shuffle a 3 carte top card (in rosso) e a riffle (in blue); accanto anche le evoluzioni rispettive a 4 carte sia top card (verde) che  riffle shuffle (nera); anche la formula approssimata viene presentata dalla linea sottile per n=4}\label{30grafico_34.pdf}
\end{center}
\end{figure}
\begin{figure}[htbp]
\begin{center}
\includegraphics[scale=0.6]{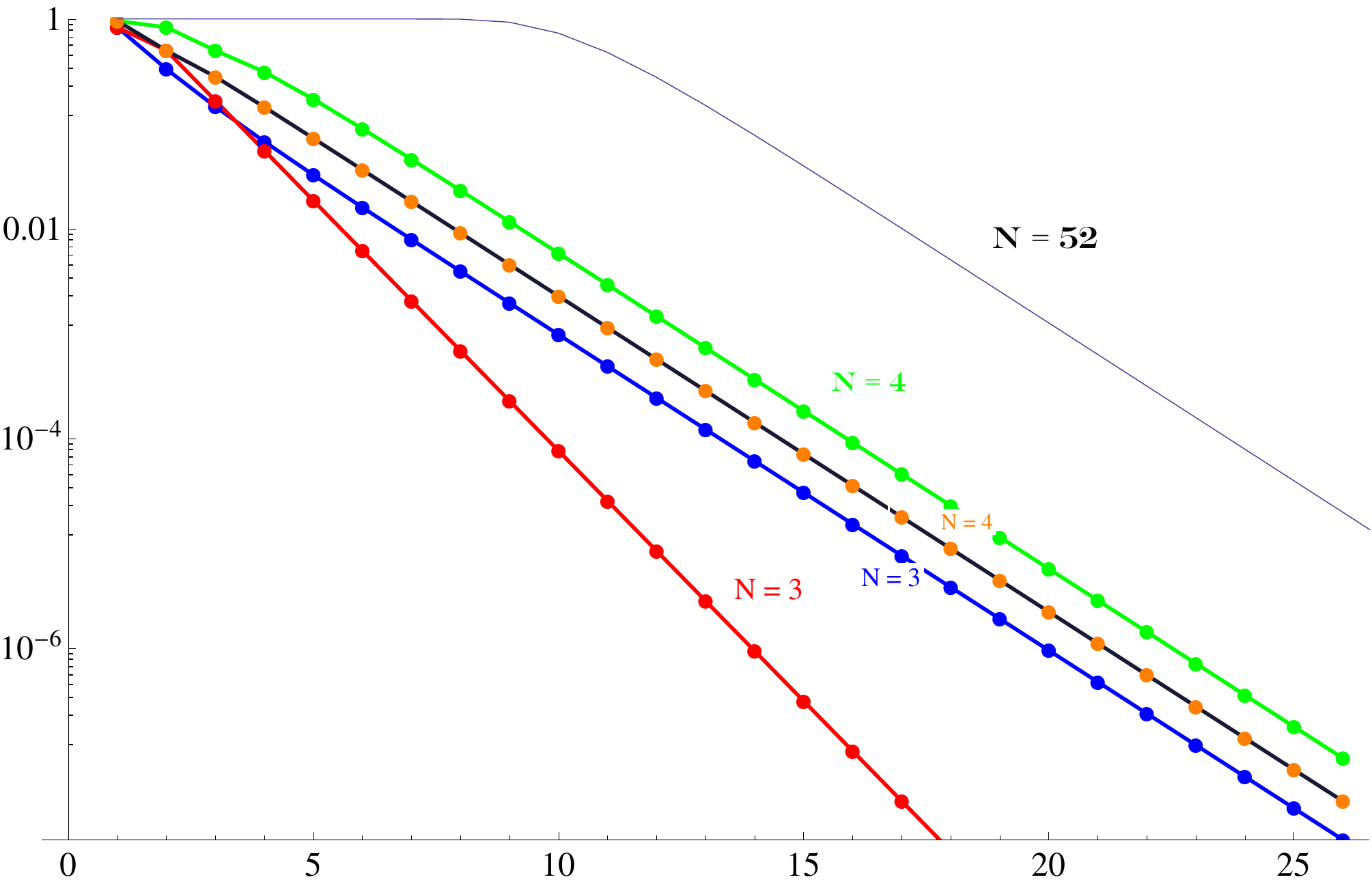}
\caption{La sequenza delle distanze in scala semilog dello shuffle a 3 carte top card (in rosso) e a riffle (in blue); accanto anche le evoluzioni rispettive a 4 carte sia top card (verde) che  riffle shuffle (nera); ancora la formula approssimata per n=52 viene descritta dalla linea sottile nera}\label{31grafico_log_etichette.pdf}
\end{center}
\end{figure}

L'andamento della distanza dalla omogeneità può essere approssimata dalla la formula approssimata
 $$ d(k) =  \|R^k - U\| \leq P(T>k) = 1 - \prod_{j=1}^{n-1}\left(1-\frac{j}{2^k}\right).$$
 e graficamente  dalla linea  continua sottile  (vedi Fig. \ref{30grafico_34.pdf} ); e
riesce a seguire bene l'andamento delle curve analitiche.
Il processo di omogenizzazione del vettore di configurazioni tramite
le matrici di transizione da noi indicate (Q) e' un processo di stati che possiamo indicare come
catena di Markov che raggiunge una perdita di memoria solo dopo un numero di passi finiti.
La  formula approssimata può essere osservata in dettaglio per 3   e 4 carte sia in top card shuffle  che in riffle shuffle mixing ottenendo in scala semilog le figure (Fig \ref{30grafico_34.pdf}, Fig.\ref{31grafico_log_etichette.pdf}) con annesso il caso della funzione approssimata in formula precedente.

In conclusione osservo come la procedura di mescolamento richieda una attenzione particolare anche nei casi più elementari, come quelli descritti. La frammentazione a pettine dei mazzetti nel riffle shuffle
induce rapidamente a serie alternate via via sempre più scarne e meno correlate tra loro.  La omogeneità, ovvero la perdita di memoria dell'ordine iniziale non si ottiene affatto in pochi mescolamenti del mazzo ma mantiene una traccia anche sottile nelle mani successive, come si evince dalle curve di decadimento d(k) in scala semilog con il mescolamento k-esimo del mazzo.
Vediamo come la mia descrizione esatta analitica-numerica e la formula approssimata più semplice in letteratura  converga e sembri ben descrivere le analoghe soluzioni analitiche a 4 carte e poi a 52 carte.
In pratica esistono tracce di memoria anche  nel mazzo apparentemente ben mischiato.
Non conosco procedure analitiche simili  a quanto  trovato in questa dissertazione.
In conclusione la informazione del mazzo è ancora attiva (o in letargo) fin tanto che vi sono grandi variazioni nei valori di probabilità  delle configurazioni da una volta alla sua successiva (cioè d(k) vicino a 1). Allorquando tali variazioni si vanno a stabilizzare allora si perde informazione: tale processo rende la distanza d(k) decrescente con legge di potenza
$2^{-k}$ che mantiene tuttavia ancora un pò una piccola traccia dell'informazione iniziale. Il problema qui analizzato con procedure numeriche elementari presenta  interessanti leggi che approssimano molto bene quelle più generali ormai note.

\section{Ringraziamenti}
Questa tesi è stata possibile grazie alla   attenzione, alla pazienza ed alle direttive del Prof. B. Scoppola
che ringrazio vivamente per tutti i numerosi suggerimenti e correzioni.
Ringrazio anche gli amici Daniele, Pietro e Linda per il loro sostegno e l'affetto.

\begin{figure}[htbp]
\begin{center}
\includegraphics[scale=0.56]{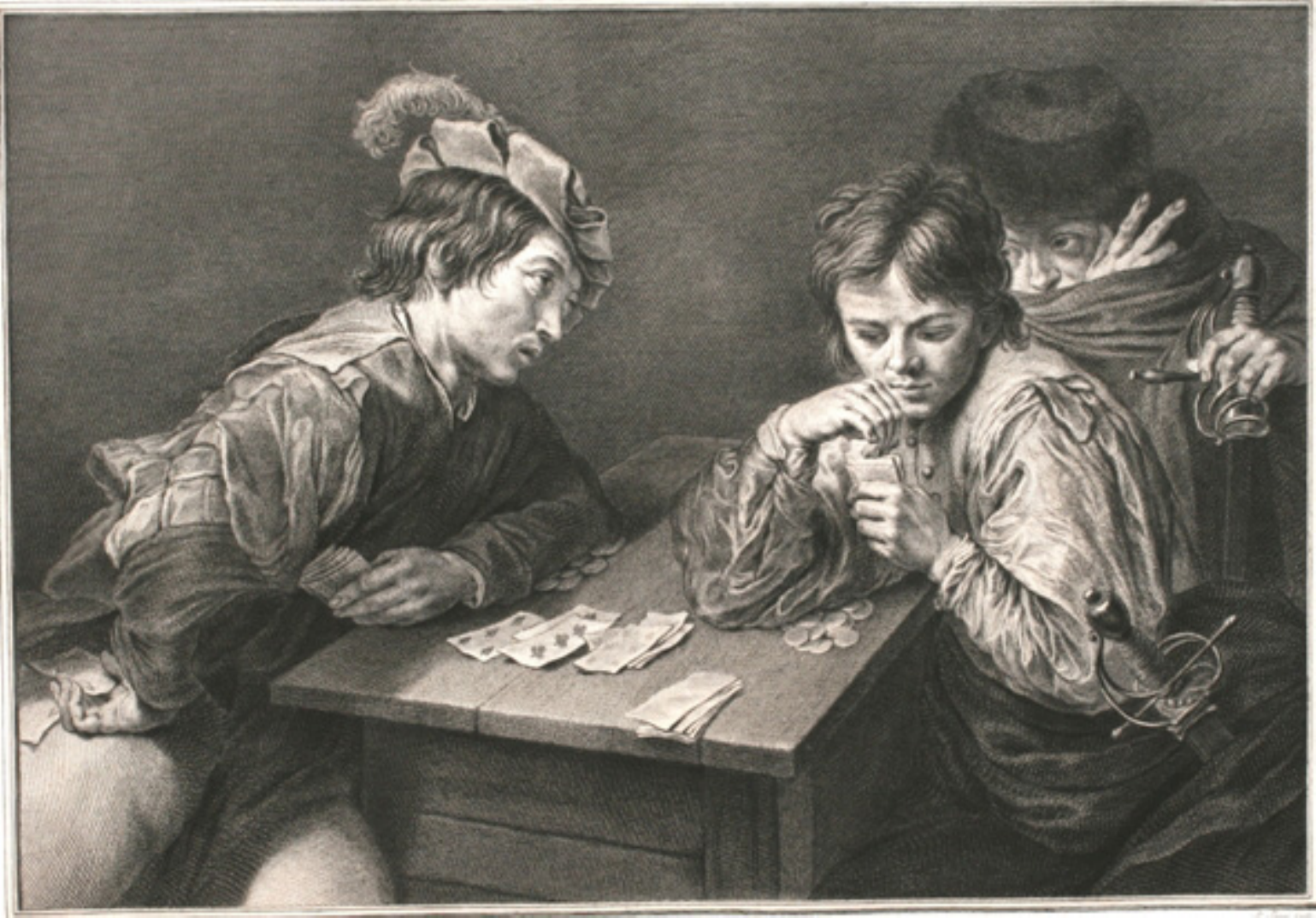}
\caption{Un quadro dell'epoca in cui le catene di Markov e la  statistica non aveva poi un gran ruolo}
\end{center}
\end{figure} 




\end{document}